\documentclass[12pt,reqno]{amsart}
\usepackage{amssymb, amsmath, hyperref}
\usepackage{pdfsync}
\usepackage{stmaryrd}
\usepackage{graphics,graphicx,fancyhdr,color}

\newtheorem{df}{Definition}[section]
\newtheorem{remark}[df]{Remark}
\newtheorem{lm}[df]{Lemma}

\newtheorem{prop}[df]{Proposition}
\newtheorem{thm}[df]{Theorem}

\makeatletter \@addtoreset{equation}{section}

\newcommand{\cal}{\mathcal}

\def\d{{\rm d}}

\newcommand{\bes}{\begin{displaymath}}
\newcommand{\ees}{\end{displaymath}}
\newcommand{\be}{\begin{equation}}
\newcommand{\ee}{\end{equation}}
\newcommand{\ba}{\begin{eqnarray}}
\newcommand{\ea}{\end{eqnarray}}
\newcommand{\bas}{\begin{eqnarray*}}
\newcommand{\eas}{\end{eqnarray*}}
\newcommand{\@Bbb}[1]{\ensuremath{\mathbb #1}}

\newcommand{\B}{{\@Bbb B}}
\newcommand{\C}{{\@Bbb C}}

\newcommand{\T}{{\mathbb T}}
\newcommand{\F}{{\@Bbb F}}
\renewcommand{\P}{{\mathbb P}}
\newcommand{\bbP}{{\P}}
\newcommand{\bbE}{{\mathbb E}}
\newcommand{\Q}{{\@Bbb Q}}
\newcommand{\bQ}{{\@Bbb Q}}
\newcommand{\N}{{\@Bbb N}}

\newcommand{\bbR}{{\mathbb R}}
\newcommand{\W}{{\@Bbb W}}
\newcommand{\Z}{{\mathbb Z}}
\newcommand{\bbZ}{{\Z}}
\newcommand{\bbT}{{\T}}

\newcommand{\la}{\lambda}
\newcommand{\al}{\alpha}
\newcommand{\ka}{\kappa}

\newcommand{\Om}{\Omega}
\newcommand{\om}{\omega}

\newcommand{\eps}{\epsilon}

\newcommand{\@s}[1]{\ensuremath{\mathcal #1}}
\newcommand{\cA}{\@s A}
\newcommand{\cB}{\@s B}
\newcommand{\cC}{\@s C}
\newcommand{\cD}{\@s D}
\newcommand{\cE}{\@s E}
\newcommand{\cF}{\@s F}
\newcommand{\cG}{\@s G}
\newcommand{\cH}{\@s H}
\newcommand{\cI}{\@s I}
\newcommand{\cJ}{\@s J}

\newcommand{\cK}{\@s K}
\newcommand{\cL}{\@s L}
\newcommand{\cN}{\@s N}
\newcommand{\cM}{\@s M}
\newcommand{\cO}{\@s O}
\newcommand{\cP}{\@s P}
\newcommand{\cR}{\@s R}
\newcommand{\cS}{\@s S}
\newcommand{\cT}{\@s T}
\newcommand{\cV}{\@s V}
\newcommand{\cW}{\@s W}
\newcommand{\cX}{\@s X}
\newcommand{\cY}{\@s Y}
\newcommand{\cZ}{\@s Z}

\newcommand{\@bm}[1]{\ensuremath{\mathbf #1}}
\newcommand{\bma}{\@bm a}
\newcommand{\bmb}{\@bm b}
\newcommand{\bmc}{\@bm c}
\newcommand{\bmd}{\@bm d}

\newcommand{\bme}{\@bm e}
\newcommand{\bmf}{\@bm f}
\newcommand{\bmg}{\@bm g}
\newcommand{\bmh}{\@bm h}
\newcommand{\bmi}{\@bm i}
\newcommand{\bmj}{\@bm j}
\newcommand{\bmk}{\@bm k}
\newcommand{\bml}{\@bm l}
\newcommand{\bmm}{\@bm m}
\newcommand{\bmn}{\@bm n}
\newcommand{\bmo}{\@bm o}
\newcommand{\bmp}{\@bm p}
\newcommand{\bmq}{\@bm q}
\newcommand{\bmr}{\@bm r}
\newcommand{\bms}{\@bm s}
\newcommand{\bmt}{\@bm t}
\newcommand{\bmu}{\@bm u}
\newcommand{\bmw}{\@bm w}
\newcommand{\bmv}{\@bm v}
\newcommand{\bmx}{\@bm x}
\newcommand{\bx}{\@bm x}

\newcommand{\bmy}{\@bm y}
\newcommand{\bz}{\@bm z}

\newcommand{\by}{\@bm y}

\newcommand{\bmzero}{\@bm 0}

\newcommand{\ga}{\gamma}

\newcommand{\@g}[1]{\ensuremath{\mathfrak #1}}
\newcommand{\gA}{\@g A}
\newcommand{\gD}{\@g D}
\newcommand{\gJ}{\@g J}
\newcommand{\gF}{\@g F}
\newcommand{\gM}{\@g M}
\newcommand{\gR}{\@g R}

\newcommand{\gq}{\@g q}
\newcommand{\gr}{\@g r}
\newcommand{\gp}{\@g p}
\newcommand{\gen}{\@g e}

\newcommand{\blambda}{\boldsymbol\lambda}

\newcommand{\commentout}[1]{{}}

\makeatother

\begin{document}

\title[Non-acoustic chain]{Diffusive propagation of energy in a
  non-acoustic chain} 

\author{Tomasz Komorowski}
\address{Tomasz Komorowski\\Institute of Mathematics, Polish Academy
  Of Sciences\\Warsaw, Poland.} 
\email{{\tt komorow@hektor.umcs.lublin.pl}}
\author{Stefano Olla}
\address{Stefano Olla\\
 CEREMADE, UMR-CNRS 7534\\
 Universit\'{e} Paris Dauphine\\
 75775 Paris CEDEX 16, France.}
 \email{{\tt olla@ceremade.dauphine.fr}}

 \begin{abstract}
{\em We consider a non acoustic chain of harmonic oscillators with the
dynamics perturbed by a random local exchange of momentum, such that
energy and momentum are conserved. The {macroscopic
  limits of 
the energy density,    momentum and the
 curvature (or bending) of the chain satisfy  a system
of evolution equations}. We prove that, in a diffusive
space-time scaling, the curvature and momentum evolve following a
linear system {that corresponds to} a
\emph{damped} {Euler-Bernoulli} beam
equation. {The macroscopic energy density} evolves
following a non linear diffusive equation. In
particular the energy transfer is diffusive in this dynamics. This
provides a first rigorous example of a normal diffusion of energy in a
one dimensional dynamics that conserves the momentum.}  
\end{abstract}
\vspace*{-1in}
\date{\today.
}
\thanks{
This paper has been partially supported by the
  European Advanced Grant {\em Macroscopic Laws and Dynamical Systems}
  (MALADY) (ERC AdG 246953), T. K. acknowledges the support of the
  Polish National Science Center grant UMO-2012/07/B/SR1/03320.}
\maketitle

\section{Introduction}
\label{intro}

Macroscopic transport in a low dimensional system, in particular the energy
transport, has attracted attention in both the physics and {mathematical
physics} literature in the latest decades. Anomalous energy
transport has been observed numerically in Fermi-Pasta-Ulam (FPU) chains,
with the diverging thermal conductivity \cite{llp97}. Generically this
anomalous superdiffusive behavior is attributed to the momentum
conservation properties of the dynamics \cite{sll}. Actually one dimensional FPU-type
chains have potential energy depending on the interparticle distances
(i.e. the gradients of the particles displacements), and have three
main locally conserved quantities: volume stretch, momentum and
energy. These conserved (or balanced) quantities have different
macroscopic space-time scalings, corresponding to different type of
initial non-equilibrium behaviour. A mechanical non-equilibrium initial
profile due to the gradients of the
  tension 
induces a macroscopic ballistic
evolution, at the hyperbolic space--time scale, governed by the Euler
equations (cf. \cite{EO}).  When the system {approaches to}, or is already at a
mechanical equilibrium, the temperature profile will evolve at a
superdiffusive time scale. 

 Recent heuristic calculations based on fluctuating hydrodynamics
 theory \cite{Sp13},  connect the macroscopic space--time scale of the
 superdiffusion of the thermal (energy) mode to the diffusive or superdiffusive
 fluctuations of the other conserved quantities. It turns out that
 this superdiffusive behavior of the energy is governed by a fractional
 laplacian heat equation. This picture can be mathematically
 rigorously proven in the case of a harmonic chain perturbed by a local random
 exchange of momentum, see  \cite{KO1,JKO}. In particular, {it has been
 shown in \cite{KO1}}, that in the
 models driven by the tension, there is a separation of the time evolution
 scales between the long modes (that evolve on a hyperbolic time scale) and
 the thermal short modes that evolve in a longer superdiffusive
 scale. In addition, from the explicit form of the
 macroscopic evolution {appearing} in these models, it is clear that this behavior
 is strongly dependent on a non-vanishing {speed} of sound. 
{More specifically}, when the speed of sound is null,
there is no {macroscopic evolution
either} at the hyperbolic or superdiffusive time scales. This suggests that the
macroscopic evolution {of the system} should happen at a yet longer, possibly diffusive, time scale
for all modes. 

In the present article we investigate the  harmonic chain model with
the random exchange of momenta. The interaction
potential depends only on the squares of the curvature (or bending)
of the chain
\begin{equation}
\label{010501}
{\frak k}_x:=-\Delta {\frak q}_x:=2 {\frak q}_{x}-{\frak q}_{x-1}-{\frak q}_{x+1},\quad
x\in\bbZ, 
\end{equation}
where ${\frak q}_{x}$ are the positions of the particles. This means
that its hamiltonian is formally given by
\begin{equation}
\label{hamilton}
{\cal H}({\frak
  k},{\frak p})=\sum_x{\frak e}_x ({\frak
  k},{\frak p}),
\end{equation}
 where the energy of the oscillator
$x$ is defined
\begin{equation}
\label{011705}
\frak e_x ({\frak k}, {\frak p}):= \frac{\frak p_x^2}{2}+ \frac{ \alpha\frak k_x^2}2 .
\end{equation}
Here $\alpha$ is a positive parameter that {indicates} the strength of the springs.
This corresponds to a special choice of attractive nearest neighbor
springs and repulsive next nearest neighbor springs. It turns out that
the {respective} speed of sound is null, even though the momentum
is conserved by the dynamics. 
As the energy depends on the curvature and not on the volume, this
system is tensionless, and the corresponding relevant conserved
quantity, besides the energy and momentum, is the
curvature and not the volume stretch.

Our first result, see Theorem \ref{profiles} below. asserts that these three conserved quantities
(curvature, momentum and energy) evolve {together} in the diffusive time scale. 
Curvature and momentum are governed macroscopically by the damped Euler-Bernoulli beam
equations:
\begin{equation}
  \label{eq:1}
  \begin{split}
    &\partial_t k(t,y) = - \Delta_y p (t,y), \\ 
&\partial_t p(t,y)
    = \alpha\Delta_y \left[k (t,y) + \vphantom{\int_0^1}3\gamma p(t,y)\right],
  \end{split}
\end{equation}
where $\gamma>0$ is the intensity of the random exchange of momentum.

Defining the mechanical macroscopic energy as
\begin{equation}
  \label{eq:mech-energy}
  e_{\text{mech}}(t,y) = \frac 12 \left( \vphantom{\int_0^1}p^2(t,y) + \alpha k^2(t,y) \right)
\end{equation}
and its thermal {counterpart} (or temperature profile) as 
\begin{equation}
  \label{eq:therm-energy}
  e_{\text{th}}(t,y) = e(t,y) - e_{\text{mech}}(t,y) 
\end{equation}
the evolution of the latter is given by
\begin{equation}
  \label{eq:5}
  \partial_t e_{\text{th}}(t,y) = \left(\frac{(\sqrt{3}-1)\alpha}{2\sqrt{3}\gamma} +
    3\gamma\right) \Delta_y  e_{\text{th}}(t,y) + 3\gamma \left(\partial_y p(t,y)\right)^2,
\end{equation}
see Theorem \ref{main-thm1}.
In particular, the thermal conductivity is finite and we have a  normal
diffusion in this system. Notice also that because of the viscosity
term, a gradient of the macroscopic velocity profile induces a local
increase of the temperature. 

This result puts in evidence two main differences {between the present and 
 the FPU-type} models:
\begin{enumerate}
\item the thermal conductivity is finite, even though the system is one
  dimensional and dynamics conserves the momentum. This {suggests} that the
  non-vanishing speed of sound is a necessary condition for
 the superdiffusion of the thermal energy, 
\item there is no separation of the time scales between low
  (mechanical) and high (thermal) {energy} modes: all the frequencies
  evolve macroscopically in the diffusive time 
  scale. Furthermore there is a continuous transfer of energy from low
  modes to high modes, {resulting in the rise} of the
  temperature, due to {the gradients of the momentum} profile. 
\end{enumerate}

These rigorous results on the harmonic non-acoustic chain {lead us}  to
conjecture that a similar behavior is expected for the deterministic
non-linear hamiltonian dynamics corresponding to an interaction of the
type $V(\frak k_x)$, i.e. {the energy is a non linear function
of the curvature of the chain.} 

About our proof of the hydrodynamic limit: this is a non-gradient
dynamics (microscopic energy currents are not of the form of discrete space
gradients of some functions). Therefore, we cannot use known techniques for
such type of limits based on relative entropy methods
(cf. e.g. \cite{varadhan-94}, \cite{Y91}) for two reasons:
\begin{enumerate}
\item lack of control of higher moments of the currents in terms of the
  relative entropy,
\item degeneracy of the noise in the dynamics, as it acts only on the velocities.
\end{enumerate}

Instead, we develop a method already used in \cite{JKO},  based on
Wigner distributions  for the energy of the acoustic chain.
Thanks to the energy conservation property of the dynamics we can
easily conclude, see Section \ref{sec5.4}, that 
the Wigner distributions  form a  compact family
of elements in a weak topology of an appropriate Banach space.
Our main result concerning the identification of its limit is contained in  Theorem \ref{main-thm} below.  
The spatial energy density is a
marginal of the Wigner function. 
We would like to highlight the fact, that
in addition
to proving the hydrodynamic limit of the energy functional, we are
also able
to identify the distribution of the macroscopic energy in the frequency
mode domain, see formula \eqref{032910}.
In particular the thermal energy is uniformly
  distributed on all modes (which is a form of \emph{local equilibrium}), while the macroscopic mechanical energy is
  concentrated on the macroscopic low modes, see \eqref{032910}.

To show Theorem \ref{main-thm}  we investigate the limit  of
the Laplace transforms of the Wigner distributions introduced in
Section \ref{sec7}. The main results, dealing with the asymptotics of
the Laplace-Wigner distributions, are formulated in Theorems \ref{cor011811} --
\ref{thm012610}.
Having these results we are able to finish the identification of the
limit of the Wigner distributions, thus
ending the proof of Theorem \ref{main-thm}.
The proofs of the aforementioned Theorems \ref{cor011811} --
\ref{thm012610}, which are rather technical,  are presented in Sections
\ref{sec4} - \ref{sec4a}, respectively.

\section{The dynamics}
\subsection{Non acoustic chain of harmonic oscillators}


Since in the non-acoustic chain the potential energy depends only on the
bendings, see \eqref{010501}, in order to describe the configuration of the
infinite chain we only need to specify $(\frak k_x)_{ x\in \mathbb
Z}$, and the configurations of our dynamics will be denoted by
 $\left((\frak p_x,\frak k_x)\right)_{ x\in \mathbb Z} \in (\mathbb R\times\mathbb
 R)^{\mathbb Z}$.


In case when no noise is present the dynamics of the chain of
oscillators  can be written formally  as  a Hamiltonian system of
differential equations  
\begin{eqnarray}
&&\dot {\frak k}_{x}(t)=-\Delta \partial_{\frak p_x}{\cal H}({\frak p}(t),{\frak k}(t))
\label{eq:bas}\\
&&\nonumber\\
&& \dot {\frak p}_x(t)= \Delta \partial_{\frak q_x}{\cal H}({\frak
   p}(t),{\frak k}(t)),\quad x\in\bbZ.\nonumber 
\end{eqnarray}
where $\Delta f(x) = f(x+1) + f(x-1) - 2f(x)$.
Let also $ \nabla g_x:=g_{x+1}-g_x$ and $ \nabla^*g_x:=g_{x-1}-g_x$.

\subsubsection{Continuous time noise}
\label{sec:cont-time-noise}


We add to the right hand side of \eqref{eq:bas} a local stochastic
term that conserves both ${\frak p}_{x-1}^2+{\frak p}_x^2+{\frak
  p}_{x+1}^2$ and ${\frak p}_{x-1}+{\frak p}_x+{\frak
  p}_{x+1}$. The respective stochastic differential equations can be
written as
\begin{eqnarray}
&d{\frak k}_x(t) &=-\Delta{\frak p}_x(t)\; dt
\label{eq:bas1},\\
&&\nonumber\\
& d{\frak p}_x(t)&=\left[\alpha\Delta{\frak k}_x(t) -
\frac{\ga}{2} \beta*{\frak p}_x(t) \right]dt \nonumber\\
&&\quad +\ga^{1/2}\sum_{z=-1,0,1} Y_{x+z}{\frak p}_x(t) dw_{x+z}(t),\quad x\in\bbZ,\nonumber
\end{eqnarray}
with the parameter $\ga>0$ that indicates  the strength of the noise
in the system, and $(Y_x)$ are vector fields given by
\begin{equation}
\label{011210}
Y_x:=({\frak p}_x-{\frak p}_{x+1})\partial_{{\frak p}_{x-1}}+({\frak p}_{x+1}-{\frak p}_{x-1})\partial_{{\frak p}_{x}}+({\frak p}_{x-1}-{\frak p}_{x})\partial_{{\frak p}_{x+1}}.
\end{equation}
Here $(w_x(t))_{t\ge0}$, $x\in\bbZ$ are i.i.d. one dimensional, real
valued, standard Brownian motions, {over a probability space} 
$(\Om,{\cal F},\bbP)$.   
Furthermore, 
 $\beta_x=\Delta\beta^{(0)}_x$, where   
$$
 \beta^{(0)}_x=\left\{
 \begin{array}{rl}
 -4,&x=0,\\
 -1,&x=\pm 1,\\
 0, &\mbox{ if otherwise.}
 \end{array}
 \right.
 $$
{As a result we obtain}
$$
 \beta_x=\left\{
 \begin{array}{rl}
 6,&x=0,\\
 -2,&x=\pm 1,\\
-1,&x=\pm2,\\
 0, &\mbox{ if otherwise.}
 \end{array}
 \right.
 $$

%

We can rewrite the system  \eqref{eq:bas1}
\begin{eqnarray}
\label{012206}
&d{\frak k}_x(t) &=-\Delta{\frak p}_x(t)\; dt
\label{eq:bas1aa},\\
&&\nonumber\\
& d{\frak p}_x(t)&=\left[\alpha \Delta\frak k_x(t)
+\frac{\ga}{2}\Delta (\beta^{(0)}*{\frak p}(t))_x \right] dt \nonumber\\
&&\quad +\ga^{1/2}\sum_{z=-1,0,1}(Y_{x+z}{\frak p}_x(t))dw_{x+z}(t),\quad x\in\bbZ.\nonumber
\end{eqnarray}

{\bf Remark.} {The particular choice of the random exchange in the above dynamics is
not important. The result can be extended to any other random mechanism
of moment exchange, as long as total energy and momentum are conserved.
Most simple dynamics would be given by exchange of momentum between nearest
neighbor atoms at independent exponential times.}

\subsection{Stationary Gibbs distributions}
\label{sec:stat-gibbs-distr}
Let $\blambda = (\beta, p, \tau)$, with $\beta^{-1}\ge 0$.
The product measures
\begin{equation}\label{eq:gibbs}
  \begin{split}
    \d\mu_{\blambda} &:= \prod_x \exp\left\{-\beta \left(\gen_x - p
        \frak p_x -\tau \frak k_x\right) - \mathcal
      G(\blambda)\right\}\; d\frak k_x
    \; d\frak p_x, \\
    \mathcal G(\blambda) &:= \frac 12 \log \left(\frac{2\pi
        \beta}{\alpha}\right) + \frac{\beta (p^2 + \alpha\tau^2)}{2}
    \end{split}
\end{equation}
are stationary for the dynamics defined by  \eqref{eq:bas1aa}. In this
context $\tau$ is called the \emph{load} of the chain, while as usual
$\beta^{-1}$ is the temperature and $p$ is the average momentum. 

Notice that, when $\tau \neq 0$, the above
  distribution is spatially translation invariant, only for the $(\frak
  k_x,\frak p_x )$ coordinates, but is is not translation invariant
  with respect to the position $\frak q_x$, or the stretch $\frak r_x
  = \frak q_x - \frak q_{x-1}$ coordinates.

\subsection{Initial data}

\label{sec:super}

Concerning the initial data we assume that, given $\eps>0$, it  is distributed according to a 
probability measure
$\mu_\eps$ on the configuration of $\left(({\frak k}_x,{\frak p}_x)\right)_{x\in\bbZ}$ and
satisfies 
\begin{equation}
\label{011605}
\sup_{\eps\in(0,1]}\eps \sum_{x}\langle {\frak e}_x \rangle_{\mu_\eps} < +\infty.
\end{equation}
Here $\langle\cdot\rangle_{\mu_\eps}$ denotes the average with respect to $\mu_\eps$.
We denote also by   $\bbE_\eps$  the expectation with respect to the
product measure
$\bbP_\eps=\mu_{\eps}\otimes \bbP$.

The existence and uniqueness of a solution to \eqref{eq:bas1} in
$\ell_2$, with the aforementioned initial condition can be easily
concluded from the standard Hilbert space theory of stochastic differential
equations, see e.g. Chapter 6 of \cite{daza}.

We assume furthermore that the mean of the initial configuration
varies on the macroscopic spatial scale:
\begin{equation}
\label{ini}
\langle{\frak k}_x\rangle_{\mu_\eps}=\ka(\eps x),\quad \langle{\frak p}_x\rangle_{\mu_\eps}=p(\eps x),\quad x\in\bbZ
\end{equation}
for some functions $\ka,p\in C_0^\infty(\bbR)$. 
Their Fourier transforms  $\hat \ka$ and $\hat p$ belong to the Schwartz class ${\cal S}(\bbR)$.
As for the fluctuations around the mean we assume that their energy
spectrum is  uniformly $L^r$ integrable with respect to $\eps>0$  for
some $r>1$.
We have denoted by
  \begin{equation}
\label{032504}
\hat f(k):=\sum_x f_x\exp\left\{-2\pi i kx\right\},\quad k\in\bbT.
\end{equation}
 the Fourier trasform of a given sequence $f_x$, $x\in\mathbb Z$.
Here $\bbT$ is the unit torus, understood  as the interval
$[-1/2,1/2]$ with the identified endpoints.
Let
\begin{equation}
\label{051708}
\tilde {\frak k}_x:={\frak k}_x-\langle{\frak k}_x\rangle_{\mu_\eps},\quad \mbox{and}\quad \tilde{\frak p}_x:={\frak p}_x-\langle{\frak p}_x\rangle_{\mu_\eps},\quad x\in\bbZ.
\end{equation}
The energy spectrum is defined as:
\begin{equation}
\label{022403}
{\cal E}_\eps(k):=\frac{1}{2}\left[\left\langle |\hat{\tilde{\frak
        p}}(k)|^2\right\rangle_{\mu_\eps}+ \alpha
     \left\langle |\hat{\tilde{\frak
      k}}(k)|^2\right\rangle_{\mu_\eps}\right],\quad k\in\bbT,
\end{equation}
where $\hat{\tilde{\frak p}}(k)$ and $\hat{\tilde{\frak k}}(k)$
are the Fourier transforms of $(\tilde{\frak p}_x)$ and $(\tilde{\frak
  k}_x)$, respectively.
Assumption \eqref{011605} implies in particular that
\begin{equation}\label{finite-energy0}
 K_0=   \sup_{\eps\in(0,1]}\eps\int_{\bbT}{\cal E}_\eps(k)dk<+\infty.
\end{equation}
The announced property of the $L^r$ integrability of the energy spectrum means  that there exists $r>1$ such that:
\begin{equation}
\label{finite-energy1}
K_1:=\sup_{\eps\in(0,1]}\eps^r\int_{\bbT}{\cal E}_{\eps}^r(k)dk<+\infty.
\end{equation}

Thanks to  the hypothesis \eqref{ini} we conclude that for any  $G\in C_0^\infty(\bbR)$  we have
\begin{eqnarray}
\label{limits0}
&&
\lim_{\eps\to0+}\eps\sum_xG(\eps x)\langle {\frak p}_x\rangle_{\mu_\eps}=\int_{\bbR}G(y)p(y)dy,\\
&&
\lim_{\eps\to0+}\eps\sum_xG(\eps x)\langle {\frak k}_x\rangle_{\mu_\eps}=\int_{\bbR}G(y)\kappa(y)dy.\nonumber
\end{eqnarray}
The quantities $p(\cdot)$,  $\kappa(\cdot)$ are called the macroscopic
{\em velocity} and {\em curvature} profiles. We assume furthermore
that the following limits exist 
\begin{eqnarray}
\label{limits2}
&&
\lim_{\eps\to0+}\eps\sum_xG(\eps x)\langle {\frak p}_x^2\rangle_{\mu_\eps}=\int_{\bbR}G(y) p_2(y)dy,\\
&&
\lim_{\eps\to0+}\eps\sum_xG(\eps x)\langle {\frak k}_x^2\rangle_{\mu_\eps}=\int_{\bbR}G(y)\kappa_2(y)dy,\nonumber\\
&&
\lim_{\eps\to0+}\eps\sum_xG(\eps x)\langle {\frak k}_x{\frak p}_x\rangle_{\mu_\eps}=\int_{\bbR}G(y)j(y)dy,\nonumber
\end{eqnarray}
for any $G\in C_0^\infty(\bbR)$. Here  $j(\cdot),p_2(\cdot)$,
$\kappa_2(\cdot)$ are some functions belonging to
$C_0^{\infty}(\bbR)$.

As a consequence, we conclude that the limit
\begin{equation}
\label{limits}
\lim_{\eps\to0+}\eps\sum_xG(\eps x)\langle {\frak e}_x\rangle_{\mu_\eps}=\int_{\bbR}G(y)e(y)dy
\end{equation}
also exists for any  $G\in C_0^\infty(\bbR)$.  Here $e(y)$ --
the {\em macroscopic energy profile} -- is given by
\begin{equation}
\label{012110}
e(y)=\frac12\left(p_2(y)+\alpha\ka_2(y)\vphantom{\int_0^1}\right).
\end{equation}
{\bf Remark.} An important example of initial distributions that satisfy the above
conditions is provided by {\em local Gibbs measures},
i.e. inhomogeneous product probability measures
of the type 
\begin{equation}\label{eq:gibbs-loc}
  \begin{split}
   \prod_{x\in\bbZ} \exp\left\{-\beta_x \left(\gen_x - p_x
        \frak p_x -\tau_x \frak k_x\right) - \mathcal
      G(\blambda_x)\right\}\; d\frak k_x
    \; d\frak p_x,
    \end{split}
\end{equation}
Here the vector $\blambda_x = \left(\beta_x, p_x, \tau_x\right)$ is
given by  $\blambda_x:= \blambda(\eps x)$, where
$\blambda(x):=\left(\beta(x), p(x), \tau(x)\right)$
and the functions $\beta^{-1}(\cdot),p(\cdot),\tau(\cdot)$ belong to
$C_0^\infty(\bbR)$.
The deterministic field $G(\blambda_x)$, $x\in\bbZ$, called the Gibbs
potential is given by an analogue of the second equality of \eqref{eq:gibbs}.

In this case $j(y) = p(y) \tau(y)$, $p_2(y) =
p(y) + \beta^{-1}(y)$, and $\kappa_2(y) = \kappa(y) + \beta^{-1}(y)$.  
{For a proof of this fact see \cite{KO1}.

\section{Formulation of the main results}
Suppose that $p(t,y)$,
$\kappa(t,y)$  satisfy the following Cauchy problem 
\begin{eqnarray}
\label{010801}
&&
\partial_t\kappa(t,y)=-\Delta_yp (t,y),\\
&&
\partial_tp(t,y)= \alpha \Delta_y \kappa(t,y)+3\gamma
\Delta_y p(t,y),\nonumber\\
&&
p(0,y)=p(y),\quad \kappa(0,y)=\kappa(y),\nonumber
\end{eqnarray}
{with $p(\cdot)$, $\kappa(\cdot)$ given by \eqref{limits0}.

Our first result concerns the evolution of the macroscopic profiles of the velocity and curvature.
\begin{thm}
\label{profiles}
Under the assumptions spelled about in the foregoing for any  $G\in C_0^\infty(\bbR)$ and $t\ge0$ we have
\begin{eqnarray}
\label{limits-t}
&&
\lim_{\eps\to0+}\eps\sum_xG(\eps x)
\bbE_\eps {\frak p}_x\left(\frac{t}{\eps^2}\right) = \int_{\bbR}G(y)p(t,y)dy,\\ 
&&
\lim_{\eps\to0+}\eps\sum_xG(\eps x)\bbE_\eps {\frak k}_x\left(\frac{t}{\eps^2}\right) =
\int_{\bbR}G(y)\kappa(t,y)dy,\nonumber
\end{eqnarray}
 where $p(t,y)$ and $\kappa(t,y)$ is the solution of \eqref{010801}.
\end{thm}
The proof of this result is fairly standard and we show it in Section \ref{sec-p-k}.
Define  the macroscopic profile of the {\em mechanical energy}
of the chain by
\begin{equation}
\label{072010}
e_{\rm mech}(t,y):=\frac12\left(p^2(t,y)+\vphantom{\int_0^1}\alpha\kappa^2(t,y)\right),
\end{equation}
with $e_{\rm mech}(y):=e_{\rm mech}(0,y)$. Comparing with the energy
profile at $t=0$, given by \eqref{012110}, we conclude that the
residual energy, called the initial {\em  thermal energy} (or temperature) profile, satisfies
\begin{equation}
\label{072010a1}
e_{\rm th}(y):=e(y)-e_{\rm mech}(y)\ge0.
\end{equation}
Concerning the evolution of the energy  profile 
we have the following result.
\begin{thm}
\label{main-thm1}
Suppose that conditions \eqref{ini}, \eqref{finite-energy1} and  \eqref{limits0} hold.
Then,   for any  $G\in C_0^\infty([0,+\infty)\times \bbR)$  the limit 
\begin{equation}
\label{energy-1}
\lim_{\eps\to0+}\eps\sum_x\int_0^{+\infty}G(t,\eps x)\bbE_\eps {\frak e}_x\left(\frac{t}{\eps^2}\right)dt=\int_0^{+\infty}\int_{\bbR}G(t,y)e(t,y)dt dy.
\end{equation}
exists. In addition,  we have
$$
e(t,y)=e_{\rm th}(t,y)+e_{\rm mech}(t,y)
$$ 
where $e_{\rm mech}(t,y)$ is given by  \eqref{072010}, while the
thermal energy (temperature) $e_{\rm th}(t,y)$ is the solution of  the following Cauchy problem:
\begin{align}
\label{energy-eqt}
&\partial_t e_{\rm th}(t,y)=\hat c\partial_y^2e_{\rm th}(t,y)+3\ga(\partial_yp)^2(t,y),\nonumber\\
&
e_{\rm th}(0,y)=e_{\rm th}(y).
\end{align}
The diffusivity coefficient equals
\begin{equation}
\label{hatc1}
  \hat c :
  = \frac{(\sqrt{3}-1)\al}{2\ga\sqrt{3}}+ 3\gamma .
  \end{equation}
\end{thm}
\bigskip

\bigskip
\begin{remark}
 {\em Notice that the gradient of the macroscopic momentum $p(t,y)$
   appearing in \eqref{energy-eqt} causes a
  local increase of the temperature. It is also straightforward to understand the
  appearance of this term in the aforementioned equation.
Consider for simplicity the case $\alpha =  0$. The dynamics is
constituted then 
 only by the random exchanges of the momentum. The conserved quantities that
evolve macroscopically are the momentum $\frak p_x$ and the kinetic
energy $\frak p_x^2/2$. The corresponding macroscopic equations
are
\begin{equation}
  \label{eq:onlyrandom}
  \begin{split}
    \partial_t p = 3\gamma \partial^2_y p \\
    \partial_t e = 3\gamma \partial^2_y e
  \end{split}
\end{equation}
These can be proven easily since the microscopic dynamics is of
gradient type. It follows that the macroscopic equation  for the temperature field, defined by
$e_{\rm th} = e - p^2/2$, 
is given by 
\begin{equation}
  \label{eq:4}
  \partial_t e_{\rm th}(t,y)=3\ga\left\{\partial_y^2e_{\rm th}(t,y) + \vphantom{\int_0^1}(\partial_yp)^2(t,y)\right\}
\end{equation}
The interaction $\alpha$ affects the thermal
diffusivity, but does not influence  the nonlinearity appearing in the evolution of
the temperature profile.
}
\end{remark}

\section{Some basic notation}

\label{sec-basic}

To abbreviate our notation we  write 
\begin{equation}
\label{021701}
\frak s(k):=\sin(\pi k)\quad \mbox{and}\quad  \frak c(k):=\cos(\pi
k),\quad k\in\bbT.
\end{equation}
Let $\ell^2 $ be the space of all complex valued sequences
$(f_x)_{x\in\bbZ}$, 
equipped with the norm $\|f\|_{\ell^2}^2:=\sum_x|f_x|^2$.
Obviously $\hat f$ belongs to $L^2(\bbT)$ - the space of
all complex valued functions equipped with the norm 
$\|f\|_{L^2(\bbT)}:=\langle \hat f,\hat f\rangle_{L^2(\bbT)}^{1/2}$, where
$$
\langle \hat f,\hat g\rangle_{L^2(\bbT)}:=\int_{\bbT}\hat
f(k)\hat g^*(k)dk,\quad \hat f,\hat g\in L^2(\bbT). 
$$


%

Given a set $A$ and two functions $f,g:A\to\mathbb R_+$ we 
say that
$
f(x)\approx
g(x)$, $x\in A$ if there exists $C>1$ such that
$$
\frac{f(x)}{C}\le g(x)\le Cf(x),\quad \forall\,x\in A.
$$
We write
$g(x)\preceq f(x)$, when only the upper bound on $g$ is satisfied.

Denote by  ${\cal S}$ the set of  functions $J:\bbR\times
\bbT\to\mathbb C$ that are of $C^\infty$ class and such that for any
integers $l,m,n$ we have 
$$
\sup_{y\in\bbR,\,k\in\bbT} (1+y^2)^{n}|\partial_y^l\partial_k^mJ(y,k)|<+\infty.
$$ 
For  $J\in {\cal S}$ we let $\hat J$ be its Fourier transform in
the first variable, i.e.
$$
\hat J(\eta,k):=\int_{\bbR}e^{-2\pi i y \eta}J(y,k)dy,\qquad
(\eta,k)\in\bbR\times \bbT.
$$

For any $M>0$ let ${\cal A}_{M}$ be the 
completion of ${\cal S}$ in the norm
\begin{equation}
\label{norm-ta0}
\| J\|_{ {\cal A}_{M}}:=\int_{B_M}d\eta\left(\int_{\bbT}|\hat J(\eta,k)|dk\right).
\end{equation}
Here $B_M:=[\eta:\,|\eta|< M]$. 
We drop the subsrcipt from the notation if $M=+\infty$.
Let ${\cal A}'$ and ${\cal A}_{M}'$ be the respective topological dual spaces of ${\cal A}$ and ${\cal A}_M$.

\section{Wigner function and its evolution}

\label{sec5}

\subsection{The wave function}

The wave function corresponding to the configuration $\left(({\frak
  p}_x,{\frak k}_x)\right)_{x\in\bbZ}$ is defined as  
\begin{equation}
\label{012010}
\psi_x:= \sqrt \alpha {\frak
  k}_x
+i{\frak p}_x,\quad x\in\bbZ.
\end{equation}
Its Fourier transform  is given by
\begin{equation}
\label{022010}
\hat\psi(k)= {\sqrt\alpha} \hat {\frak
  k}\left(k\right)+i\hat{\frak p}\left(k\right),\quad k\in\bbT. 
\end{equation}
The energy and its spectrum \eqref{022403} can be written as
\begin{equation}
\label{052010}
{\frak e}_x=\frac12 |\psi_x|^2,\quad x\in\bbZ\quad \mbox{and}\quad {\cal E}_\eps(k)=\frac{1}{2}\langle|\hat\psi(k)|^2\rangle_{\mu_\eps},\quad k\in\bbT.
\end{equation}
Using the decomposition into the macroscopic profile and the fluctuation part, see \eqref{051708}, we can  write
  \begin{equation}
\label{032010}
\psi_x= \phi(\eps x)+\tilde\psi^{(\eps)}_x,\quad x\in\bbZ,
\end{equation}
where
$$
\phi(y):={\sqrt\alpha}\ka(y) + ip(y)\quad \mbox{ and }
\quad \tilde\psi^{(\eps)}_x:={\sqrt\alpha}\tilde{\frak k}_x + i\tilde {\frak p}_x
$$ 
are the wave functions corresponding to the macroscopic profile and  the fluctuation part, respectively.

\subsection{Wigner functions}

By the Wigner  functions corresponding to  the wave function field $\left(\psi_x\right)_{x\in\bbZ}$ we understand four tempered distributions 
$W_{\eps,\pm}$, $Y_{\eps,\pm}$ that we often write together in the form of a  vector  
\begin{equation}
\label{032610-1}
 {\frak W}_\eps^T:=[W_{\eps,+},Y_{\eps,+},Y_{\eps,-},W_{\eps,-}],
 \end{equation}
where
 \begin{equation}
 \label{wigner1}
\langle
W_{\eps,\pm},J\rangle:= \int_{\bbR\times\bbT}
 \widehat W_{\eps,\pm}(\eta,k)\hat  J^*(\eta, k)d\eta dk,
\end{equation}
and
\begin{equation}
 \label{wigner2}
\langle Y_{\eps,\pm},
J\rangle:= \int_{\bbR\times\bbT} \widehat Y_{\eps,\pm}(\eta,k)\hat J^*(\eta,k)d\eta dk
\end{equation}
for any $ J\in {\cal A}$. Here   $\widehat W_{\eps,\pm}(\eta,k)$ and $\widehat Y_{\eps,\pm}(\eta,k)$ -- called {\em the Fourier-Wigner} functions  -- are given by
 \begin{eqnarray}
 \label{021507}
 &&
 \widehat W_{\eps,+}(\eta,k):=\frac{\eps}{2}\left\langle\left(\hat \psi\right)^*\left(k-\frac{\eps \eta}{2}\right)\hat \psi\left(k+\frac{\eps \eta}{2}\right)\right\rangle_{\mu_\eps},\nonumber\\
&&
\widehat Y_{\eps,+}(\eta,k):=\frac{\eps}{2}\left\langle\hat \psi\left(- k+\frac{\eps \eta}{2}\right)\hat \psi\left( k+\frac{\eps \eta}{2}\right)\right\rangle_{\mu_\eps}, \\
&&\widehat Y_{\eps,-}(\eta,k) :=\widehat Y_{\eps,+}^*(-\eta,k),\quad
\widehat W_{\eps,-}(\eta,k) :=\widehat W_{\eps,+}(\eta,-k). \nonumber
\end{eqnarray}
For any $ J\in {\cal A}$ we can write
$$
|\langle W_{\eps,+},J\rangle|\le \frac{\eps}{2}\| J\|_{\cal A}\sup_{\eta}\int_{\bbT}
\left|\left\langle \hat \psi^{(\eps)}\left(k-\frac{\eps \eta}{2}\right)\hat
  \psi^{(\eps)}\left(k+\frac{\eps \eta}{2}\right)\right\rangle_{\mu_\eps}\right|dk.
$$
Using the Cauchy-Schwartz inequality  and \eqref{finite-energy0} we  get
   \begin{equation}
 \label{A1}
\sup_{\eps>0}\sum_{\iota=\pm}(\| Y_{\eps,\iota}\|_{{\cal A}'}+\|
W_{\eps,\iota}\|_{{\cal A}'})\le 4K_0.
\end{equation}

A simple calculation shows that for any function $J(y,k)\equiv J(y)$ 
\begin{equation}
  \label{eq:waveenergy}
  \langle W_{\eps,\pm}, J \rangle = \eps \sum_x 
  \langle{\frak e}_x\rangle_{\mu_\eps}J^*(\pm\eps x)
\end{equation}
and
\begin{equation}
  \label{eq:waveenergy1}
  \langle Y_{\eps,\pm}, J \rangle = \eps \sum_x 
  \langle{\frak l}_x\pm i \sqrt \alpha{\frak j}_x\rangle_{\mu_\eps}J^*(\eps x),
\end{equation}
where
$$
{\frak l}_x:= \frac12\left(\alpha{\frak k}_x^2-
{\frak p}_x^2\right),\qquad {\frak j}_x:={\frak k}_x{\frak p}_x,\quad x\in\bbZ. 
$$

Using the decomposition of the wave function into its mean, following a macroscopic profile $\phi(\cdot)$, and the fluctuation part $\{\tilde\psi_x^{(\eps)},\,x\in\bbZ\}$, see \eqref{032010},
we can correspondingly decompose the vector of the Wigner functions. Namely 
\begin{equation}
\label{decomp}
 {\frak W}_\eps =\overline{\frak W}_\eps+\widetilde{\frak W}_\eps,
\end{equation}
where the Fourier-Wigner function corresponding to these wave functions shall be denoted by
$$
 \overline{\frak W}_\eps^T:=[\overline{W}_{\eps,+},\overline{Y}_{\eps,+},\overline{Y}_{\eps,-},\overline{W}_{\eps,-}].
 $$
 and 
 $$
 \widetilde{\frak W}_\eps^T:=[ \widetilde{W}_{\eps,+}, \widetilde{Y}_{\eps,+}, \widetilde{Y}_{\eps,-}, \widetilde{W}_{\eps,-}].
 $$
We let
\begin{eqnarray*}
&&
\langle \overline{W}_{\eps,\pm},J\rangle:=\int_{\bbR\times\bbT}\widehat{\overline{
    W}}_{\eps,\pm}(\eta,k)J^*(\eta,k)d\eta dk,\\
    &&
     \langle {\widetilde{
    W}}_{\eps,\pm},J\rangle:=\int_{\bbR\times\bbT}\widehat{\widetilde{
    W}}_{\eps,\pm}(\eta,k)J^*(\eta,k)d\eta dk,
\end{eqnarray*}
where, using the Poisson summation formula, we have defined
\begin{eqnarray}
\label{041308}
&&
\widehat{\overline{ W}}_{\eps,\pm}(\eta,k)=\frac{1}{2\eps}\sum_{x,x'}\hat\phi^*\left(\frac{\pm k+x}{\eps}-\frac{\eta}{2}\right)\hat\phi\left(\frac{\pm k+x'}{\eps}+\frac{\eta}{2}\right),\nonumber\\
&&
\widehat{\widetilde{
    W}}_{\eps,\pm}(\eta,k)=\frac{\eps}{2} \left\langle\left(\hat {\tilde\psi}^{(\eps)}\right)^*\left(\pm k-\frac{\eps \eta}{2}\right)\hat
 { \tilde\psi}^{(\eps)}\left(\pm k+\frac{\eps \eta}{2}\right)\right\rangle_{\mu_\eps}.
\end{eqnarray}
The formulas for $\langle \overline{Y}_{\eps,\pm},J\rangle$ and  
$\langle {\widetilde{ Y}}_{\eps,\pm},J\rangle$ are constructed
analogously using  the respective Fourier-Wigner functions. 
Notice that for small $\eps$ the expression above of
$\widehat{\overline{ W}}_{\eps,\pm}$ is well approximated by the more
natural definition:
$$
\widehat{\overline{W}}_{\eps,\pm}(\eta,k) \sim
   \frac{\eps}{2}\hat\phi^*\left(\frac{\pm k}{\eps}-\frac{\eta}{2}\right) 
\hat\phi\left(\frac{\pm k}{\eps}+\frac{\eta}{2}\right).
$$

As a consequence of assumption
\eqref{limits2}
we conclude that for functions $J(y,k) = J(y)$:
\begin{eqnarray*}
&&
\lim_{\eps\to0+}\langle W_{\eps,\pm}, J \rangle=\int_{\bbR}e(\pm y)J^*(y)dy,
\\
&&
\lim_{\eps\to0+}\langle Y_{\eps,\pm}, J \rangle=\int_{\bbR}\left(l(y)\pm i\sqrt\alpha j(y)\right)J^*(y)dy
\end{eqnarray*}
and 
$$
 l(y):=\frac12 \left(\alpha\ka_2(y)- p_2(y)\right),
$$
with $j(\cdot)$, $\ka_2(\cdot)$ and $p_2(\cdot)$ given by \eqref{limits2}.
A simple calculation also shows that
\begin{equation}
\label{020801}
\begin{split}
  \lim_{\eps\to0+} \overline W_{\eps,\pm}(J) =\overline
  W_\pm(J):=\frac12\int_{\bbR^2}
  \hat\phi^*\left(h-\frac{\eta}{2}\right) \hat\phi\left(h
    +\frac{\eta}{2}\right)\hat J^*(\eta,0)d\eta dh\\
  = \frac12\int_{\bbR^2} 
  |\phi^*\left(y\right)|^2  e^{-2\pi i y \eta}\hat J^*(\eta,0)d\eta dy
\end{split}
\end{equation}
Thus,
\begin{equation}
\label{082010}
\overline W_\pm(dy,dk)=\frac12|\phi(y)|^2\delta_0(dk)dy.
\end{equation}
One can also easily check that
\begin{equation}
\label{092010}
\overline Y_+(dy,dk)=\frac12\phi^2(y)\delta_0(dk)dy\quad\mbox{and}\quad \overline Y_-(dy,dk)=\frac12[\phi^*(y)]^2\delta_0(dk)dy.
\end{equation}
  We denote the respective vector
$
 \overline{\frak W}^T:=[\overline{W}_{+},\overline{Y}_{+},\overline{Y}_{-},\overline{W}_{+}].
 $

\subsection{Evolution of the wave function}

Adjusted to the macroscopic time, we can define the wave function corresponding to the configuration at  time $t/\eps^2$
\begin{equation}
\label{011307}
\psi^{(\eps)}_x(t):= \sqrt \alpha{\frak
  k}_x\left(\eps^{-2} t\right) 
+i{\frak p}_x\left(\eps^{-2} t\right),\quad x\in\bbZ,
\end{equation}
where  $({\frak p}_x(t),{\frak k}_x(t))_{x\in\bbZ}$ satisfies \eqref{eq:bas1}. 
Its Fourier transform
\begin{equation}
\label{011307a}
\hat\psi^{(\eps)}(t,k)=\sqrt\alpha \hat {\frak k}\left(\frac{t}{\eps^{2}},k\right)+i\hat{\frak p}\left(\frac{t}{\eps^{2}},k\right),\quad k\in\bbT,
\end{equation}
is the unique  solution of the It\^o stochastic
differential equation, understood in the mild sense (see e.g. Theorem  7.4 of \cite{daza})
  \begin{eqnarray}
 \label{basic:sde:2}
 &&
 d\hat\psi^{(\eps)}(t,k)=\left\{\frac{-i\om(k)}{\eps^{2}}\hat\psi^{(\eps)}(t,k)-\frac{\ga
   R(k)}{\eps^{2}}\left[\hat\psi^{(\eps)}(t,k)-(\hat\psi^{(\eps)})^*(t,-k)\right]\right\}dt\nonumber\\
 &&
 \\
 &&
  +\frac{i\ga^{1/2}}{\eps}\int_{\bbT}
  r(k,k')\left[\hat\psi^{(\eps)}(t,k-k')-(\hat\psi^{(\eps)})^*(t,k'-k)\right]B(dt,dk'),\nonumber
 \end{eqnarray}
 where $ \hat\psi^{(\eps)}(0)\in L^2(\bbT)$,
\begin{equation}
\label{om2}
\om(k) := 2\sqrt\alpha {\frak s}^2(k),\quad k\in\bbT
\end{equation}
is a {\em dispersion relation},
\begin{equation}
\label{beta}
R(k):=2\frak s^2( k)\left[1+2\frak c^2(k)\right]=2\frak s^2(
k)+4\frak s^2(2 k).
\end{equation}
and 
 \begin{eqnarray}
 \label{r}
&& r(k,k'):= 
4{\frak s}(k){\frak s} (k-k'){\frak s}(2k-k'),\quad k,k'\in \bbT.
 \end{eqnarray}
 A simple calculation shows that $\hat \beta(k)=4R(k)$.

The process $B(dt,dk)$ is a space-time Gaussian white noise, 
i.e.
$$
 \bbE\left[B(dt,dk)B^*(ds,dk')\right]=\delta(t-s)\delta(k-k')dtds dk dk'.
$$
Since the total energy of the system is conserved in time, see Section 2 of \cite{BOS}, 
for each $\eps\in(0,1]$ we have
\begin{equation}
\label{psi-as}
\|\hat \psi^{(\eps)}(t)\|_{L^2(\bbT)}=\|\hat \psi^{(\eps)}\|_{L^2(\bbT)},\qquad t\ge0,\quad \bbP_\eps \mbox{ a.s.}
\end{equation}

\subsection{Wigner functions corresponding to $\psi^{(\eps)}(t)$}
\label{sec5.4}

Denote by 
$$
 {\frak W}_\eps^T(t):=[W_{\eps,+}(t),Y_{\eps,+}(t),Y_{\eps,-}(t),W_{\eps,-}(t)]
 $$
the vector made of Wigner functions corresponding to the wave functions $\psi^{(\eps)}(t)$. 
They can be defined by formulas \eqref{wigner1} and \eqref{wigner2}, where the respective Fourier-Wigner functions 
$\widehat W_{\eps,\pm}(t,\eta,k)$ and $\widehat Y_{\eps,\pm}(t,\eta,k)$
 are given by analogues of \eqref{021507}
 in which the wave functions are substituted by $\psi^{(\eps)}(t)$ and the average $\langle \cdot\rangle_{\mu_\eps}$ is replaced by 
 $\bbE_\eps$.



From \eqref{psi-as}
we
conclude, thanks to \eqref{A1},  that
\begin{equation}
\label{W-Y}
\sup_{t\ge0}\sum_{\iota=\pm}\left(\sup_{\eps\in(0,1]}\| W_{\eps,\iota}(t)\|_{{\cal A}'}+\sup_{\eps\in(0,1]}\| Y_{\eps,\iota}(t)\|_{{\cal A}'}\right)\le 4K_0,
\end{equation}
where $K_0$ is the constant appearing in condition \eqref{finite-energy0}.
As a direct consequence of the above estimate we infer that the components of  $\left({\frak W}_\eps(\cdot)\right)_{\eps\in(0,1]}$ are 
 $*-$weakly sequentially compact in 
 $\left(L^1([0,+\infty);{\cal A})\right)^*$ as $\eps\to0+$, i.e. given a component of the above family, e.g.  $W_{\eps,+}(\cdot)$,  and any sequence $\eps_n\to0+$ one can choose a subsequence $W_{\eps_{n'},+}(\cdot)$ 
 converging $*$-weakly.

To characterize the limit we recall that the thermal energy density $e_{\rm th}(t,y)$ is given by the solution of the Cauchy problem \eqref{energy-eqt},
while the mechanical one $e_{\rm mech}(t,y)$
is defined by \eqref{072010}.
The limit of the Wigner functions corresponding to the
macroscopic profile wave function 
\begin{equation}
\label{012810}
\phi(t,y):=\sqrt\alpha \kappa(t,y)+ip(t,y),\quad (t,y)\in[0,+\infty)\times \bbR.
\end{equation}
equals
\begin{equation}
\label{bW}
 \overline{\frak W}^T(t):=[\overline{W}_{+}(t),\overline{Y}_{+}(t),\overline{Y}_{-}(t),\overline{W}_{+}(t)],
 \end{equation}
where
\begin{equation}
\label{082010a}
\overline W_\pm(t,dy,dk)=\frac12|\phi(t,y)|^2dy\delta_0(dk)=e_{\rm mech}(t,y)dy\delta_0(dk).
\end{equation}
and $\overline Y_-(t)=\overline Y_+^*(t)$, with
\begin{eqnarray}
\label{092010a}
&&
\overline Y_+(t,dy,dk)=\frac12\phi^2(t,y)dy\delta_0(dk).
\end{eqnarray}

Our main result concerning the limit of the Wigner transform can be stated as follows.
\begin{thm}
\label{main-thm}
Suppose that the initial data satisfy the assumptions Then,  $({\frak W}_\eps(t))_{t\ge0}$ converge, as $\eps\to0+$, $*$-weakly over 
$\left(L^1([0,+\infty),{\cal A})\right)^*$ to 
\begin{equation}
\label{013108}
{\frak W}^T(t)=[W(t),\overline Y_+(t),\overline Y_-(t),W(t)],\quad t\ge0,
\end{equation}
where $
 W(t)
$
is a measure on $\bbR\times\bbT$ given by
\begin{equation}
\label{032910}
 W(t,dy,dk):=e_{\rm th}(t,y)dydk+e_{\rm mech}(t,y)dy\delta(dk).
\end{equation}
\end{thm}

\bigskip

Analogously to  formulas  \eqref{eq:waveenergy} and \eqref{eq:waveenergy1}  we can write 
\begin{equation}
\label{eq:waveenergy-a}
\eps\sum_x\bbE_\eps {\frak e}_x\left(\frac{t}{\eps^2}\right)J^*(\pm\eps x)=\langle W_{\eps,\pm}(t),J\rangle
\end{equation}
and 
\begin{equation}
\label{eq:waveenergy-1a}
\eps\sum_x\bbE_\eps\left[ {\frak l}_x\left(\frac{t}{\eps^2}\right)\pm
  i\sqrt\alpha  {\frak j}_x\left(\frac{t}{\eps^2}\right)\right]J^*(\eps x)
= \langle Y_{\eps,\pm}(t),J\rangle,\quad J\in {\cal S}(\bbR).
\end{equation}
Therefore, the conclusion of Theorem \ref{main-thm1} is a direct consequence of Theorem \ref{main-thm}.

\section{Evolution of the Wigner functions}

\label{sec6}

Using \eqref{basic:sde:2} we can derive the equations describing the time evolution of the Wigner functions.
In particular,  one can conclude that for a fixed $\eps$ the components of $({\frak W}_\eps(t) )_{t\ge0}$ belong to $C([0,+\infty);{\cal A}')$.
After  a straightforward calculation (see Section 8 of \cite{JKO} for
details) we obtain that their Fourier transforms satisfy
\begin{equation}
\label{exp-wigner-eqt-1}
\partial_t\widehat W_{\eps,+}(t)=-\frac{i}{\eps}\delta_{\eps}\om
\widehat W_{\eps,+}(t)
+\frac{\ga}{\eps^{2}}{\cal L}_{\eps \eta }\widehat W_{\eps,+}(t)
-\frac{\ga}{2\eps^{2}}\sum_{\iota=\pm}{\cal L}^{+}_{\iota\eps \eta }\widehat Y_{\eps,-\iota}(t),
\end{equation}
and
\begin{eqnarray}
\label{anti-wigner-eqt1}
\partial_t\widehat Y_{\eps,+}(t)&=&-\frac{2i}{\eps^{2}} \bar\om\widehat Y_{\eps,+}(t)+\frac{\ga}{\eps^{2}}  {\cal L}_{\eps \eta }\widehat Y_{\eps,+}(t) \qquad~~~~~~~~~~~\\
&&
+\frac{\ga}{\eps^{2}} {\cal R}_{\eps \eta }(\widehat
Y_{\eps,-}-\widehat Y_{\eps,+} )(t) 
-\frac{\ga}{2\eps^{2}}\sum_{\iota=\pm} {\cal L}^+_{\iota\eps \eta }
\widehat W_{\eps,-\iota}(t) . \nonumber 
\end{eqnarray}
Here (cf \eqref{om2})
\begin{eqnarray}
\label{050411}
&&
\delta_{\eps}\om:=\frac{1}{\eps}\left[\om\left(k+\frac{\eps
      \eta}{2}\right)-\om\left(k-\frac{\eps
      \eta}{2}\right)\right]= 2\sqrt\alpha{\frak s}(\eps \eta){\frak s}(2k),\nonumber\\
&&
\bar \om:=\frac{1}{2}\left[\om\left(k+\frac{\eps
   \eta}{2}\right)+\om\left(k-\frac{\eps \eta}{2}\right)\right] 
   =4\sqrt\alpha \left[{\frak s}^2\left(k\right){\frak c}^2\left(\frac{\eps
      \eta}{2}\right)+{\frak c}^2\left(k\right){\frak s}^2\left(\frac{\eps
      \eta}{2}\right)\right],\nonumber\\
&&
{\cal L}_\eta  f(k):=2{\cal R}_{ \eta } f(k) - 2 f(k) \int_{\bbT} R(k,k',\eta )dk',\nonumber\\
&&
\\
&&
{\cal L}_{\eta }^\pm f(k):=2{\cal R}_{\eta } f(k) -2R\left(k\pm\frac \eta 2\right)
f(k),\nonumber \\
&&
 {\cal R}_\eta  f(k):=\int_{\bbT} R(k,k',\eta )f(k')dk',\nonumber
\end{eqnarray}
where  the scattering kernel  (cf \eqref{r})
\begin{equation}
\label{R}
R(k,k',\eta ):=\frac12\sum_{\iota=\pm1}r\left(k-\frac{ \eta }{2},k-\iota k'\right)r\left(k+\frac{ \eta }{2},k-\iota k'\right),\quad k,k'\in\bbT.
\end{equation}

A direct calculation yields
\begin{equation}
\label{022110}
R(k,k',\eta )
= R(k,k')-{\frak s}^2\left( \frac{ \eta }{2}\right) R_1(k,k')+{\frak s}^4\left( \frac{ \eta }{2}\right) R_2(k,k';\eta).
\end{equation}
Here  
\begin{eqnarray*}
&&
R(k,k'):=R(k,k', 0)=\dfrac{3}{4}\left({\frak e}_-\otimes {\frak e}_+\vphantom{\int_0^1}+{\frak e}_+\otimes {\frak e}_-\right)(k,k'),\\
&&
R_1(k,k')
:=\left(16\frak f_+\otimes \frak f_{+}+\frak f_+\otimes \frak e_{-}\vphantom{\int_0^1}+\frak e_{-}\otimes\frak f_+ +3\frak f_{-}\otimes  \frak e_{+} +3 \frak e_{+}\otimes\frak f_{-}\right)(k,k') ,
\\
&&
R_2(k,k';\eta)
=16\left(\frak{f}_+(k)
   \vphantom{\int_0^1}+\frak{f}_+(k')\right)+4\left(4\frak{f}_+\otimes\frak{f}_++
   \vphantom{\int_0^1} \frak{f}_+\otimes\frak{f}_-+\frak{f}_-\otimes\frak{f}_+\right)(k,k')\\
&&
-32{\frak
  s}^2\left( \frac{ \eta
  }{2}\right)\left(\frak{f}_+(k)+\frak{f}_+(k') \vphantom{\int_0^1}+2{\frak f}(k)\right)+64{\frak
  s}^4\left( \frac{ \eta
  }{2}\right){\frak f}(k)
,
\end{eqnarray*}
where $\frak{e}_\pm$ and $\frak{f}_\pm$ are the $L^1(\bbT)$ normalized vectors given by
\begin{equation}
\label{frak-e-r}
\frak{e}_+(k):=\frac{8}{3}{\frak s}^4( k),\quad
\frak{e}_{-}(k):=2{\frak s}^2(2 k)
\end{equation}
and ${\frak f}\equiv1$,
\begin{equation}
\label{011701a}
\frak{f}_+(k):=2{\frak s}^2( k),\quad \frak{f}_{-}(k):=2{\frak c}^2(
k),\quad k\in\bbT.
\end{equation}
Note also that (cf \eqref{beta})
\begin{equation}
 \label{072110}
 R(k)=\int_{\bbT} R(k,k')dk'=\frac{3}{4}\sum_{\iota\in\{-,+\}} \frak e_\iota(k).
\end{equation}
In addition
\begin{equation}
\label{beta1}
R'(k)=2\pi ({\frak s}(2 k)+{\frak s}(4 k))
\end{equation}
and
\begin{equation}
\label{beta2}
R''(k)=
4\pi^2(4{\frak c}^2(2 k)+{\frak c}(2 k)-2).
\end{equation}

\subsection{System of equations for the Laplace-Fourier transform of the Wigner functions}

Taking the Laplace transform of both sides of
\eqref{exp-wigner-eqt-1} and
\eqref{anti-wigner-eqt1}, we get the following equations
\begin{equation}
\label{020911a}
\tilde D_{1}^{(\eps)} w_{\eps,+}+\tilde D_{+}^{(\eps)} y_{\eps,+} +\tilde D_{-}^{(\eps)}  y_{\eps,-}
={\cal R}_1^{(\eps)}
\end{equation}
and
\begin{equation}
\label{040911a}
\tilde D_{+}^{(\eps)} w_{\eps,+}+\tilde D_{2}^{(\eps)} y_{\eps,+} +\tilde D_{-}^{(\eps)}
w_{\eps,-}={\cal R}_2^{(\eps)}.
\end{equation}
Here
\begin{eqnarray*}
&&
 w_{\eps,\pm}(\la,\eta,k):=\int_0^{+\infty}e^{-\la t}\widehat W_{\eps,\pm}(t,\eta,k)dt,\\
&&
 y_{\eps,\pm}(\la,\eta,k):=\int_0^{+\infty}e^{-\la t}\widehat Y_{\eps,\pm}(t,\eta,k)dt.
\end{eqnarray*}
In addition, we let
\begin{eqnarray}
\label{030911}
&&
\tilde D_{1}^{(\eps)}(\la,\eta,k):= \eps^{2}\la+2\ga
R_\eps+i\eps\delta_\eps\om\\
&&
\tilde D_{2}^{(\eps)}(\la,\eta,k):=\eps^{2}\la+2\ga
R_\eps+2i\bar\om,\nonumber\\
&&
\tilde  D_{\pm}^{(\eps)}(\la,\eta,k):=-\ga R_\eps \pm\dfrac{\ga \eps}{2}
  R'\eta,\nonumber
\end{eqnarray}
where
\begin{equation}
\label{r-eps}
R_\eps:=R+\frac{(\eps
      \eta)^2}{8}R''.
\end{equation}

The right hand sides of \eqref{020911a} and \eqref{040911a}  are respectively equal
\begin{eqnarray}
\label{rhs}
&&
{\cal R}_1^{(\eps)}:=g_\eps+\eps^{2}\widehat W_{\eps,+}(\eta,k)-
\frac{\ga (\pi\eps \eta)^2}{2} f_\eps +\eps^{3}
r_\eps^{(1)},\nonumber\\
&&
\\
&&
{\cal R}_2^{(\eps)}:=-g_\eps+\eps^{2}\widehat Y_{\eps,+}(\eta,k)+
\frac{\ga (\pi\eps \eta)^2}{2} f_\eps +\eps^{3}
r_\eps^{(2)},\nonumber
\end{eqnarray}
where
\begin{eqnarray}
\label{f-eps}
&&
g_\eps:=\frac 32\ga\sum_{\iota=\pm}\frak
e_{\iota}\langle v_\eps,\frak
e_{-\iota}\rangle_{L^2(\bbT)}
,\\
&&
 f_\eps :=\frak
  f_+\langle v_{\eps},16\frak f_++\frak e_{-}\rangle_{L^2(\bbT)}+\frak
  e_{-}\langle v_{\eps},\frak
  f_{+}\rangle_{L^2(\bbT)}+3\frak
  f_{-}\langle v_{\eps},\frak
  e_{+}\rangle_{L^2(\bbT)}+3\frak
  e_{+}\langle v_{\eps},\frak
  f_{-}\rangle_{L^2(\bbT)}.\nonumber
\end{eqnarray}
Here, for the abbreviation sake we have let
\begin{equation}
\label{v}
v_\eps(\la,\eta ,k):=
  w_{\eps,+}(\la,\eta ,k)- \frac12
  y_{\eps,o}(\la,\eta ,k),
\end{equation}
where
$$
 y_{\eps,o}(\la,\eta ,k):= y_{\eps,+}(\la,\eta ,k)+ y_{\eps,-}(\la,\eta ,k).
$$
In addition,  the remainder terms $ r_\eps^{(i)}$, $i=1,2$ 
 satisfy 
\begin{equation}
 \label{f}
 \limsup_{\eps\to0+}\sup_{\la\ge\la_0}\| r^{(i)}_\eps(\la)\|_{{\cal A}'_M}<+\infty,\quad i=1,2,\,\la_0,M>0.
 \end{equation}

A closed system of equations on 
\begin{equation}
\label{w}
{\frak w}_\eps^T(\la,\eta ,k):=\left[
 w_{\eps,+},
 y_{\eps,+},
 y_{\eps,-},
 w_{\eps,-}
 \right]
\end{equation}
can be rewritten in  the matrix form
\begin{equation}
\label{011812}
\tilde D_\eps{\frak w}_\eps=\frak R_{\eps},
\end{equation}
where
$$
\frak R_{\eps}^T:=\left[
 {\cal R}_1^{(\eps)}, {\cal R}_2^{(\eps)}, {\cal R}_{2,-}^{(\eps)}, {\cal R}_{1,-}^{(\eps)}\right]
$$
and $\tilde D_\eps$ is a $4\times 4$ matrix that can be written in the
block form
\begin{equation}
 \label{011612}
 \tilde D_\eps=
 \left[
 \begin{array}{cc}
A_\eps &B_\eps \\
B_\eps&C_\eps
 \end{array}
 \right],
  \end{equation}
where $A_\eps$, $B_\eps$, $C_\eps$ are $2\times 2$ matrices given by 
\begin{equation}
 \label{012410}
 A_\eps:=
 \left[
 \begin{array}{cc}
\tilde D_{1}^{(\eps)}& \tilde D_{+}^{(\eps)}  \\
\tilde D_{+}^{(\eps)}& \tilde D_{2}^{(\eps)}
 \end{array}
 \right],\qquad C_\eps:=
 \left[
 \begin{array}{cc}
 \left(\tilde D_{2}^{(\eps)}\right)^* &\tilde D_{+}^{(\eps)} \\
\tilde D_{+}^{(\eps)} &\left(\tilde D_{1}^{(\eps)}\right)^*
 \end{array}
 \right]
  \end{equation}
and $B_\eps=\tilde D_{-}^{(\eps)}I_2$,  with $I_n$ denoting the
$n\times n$ identity matrix.

We have also denoted
$$
{\cal R}_{1,-}^{(\eps)}(\la,\eta ,k):={\cal
    R}_{1}^{(\eps)}(\la,\eta ,-k),\qquad{\cal R}_{2,-}^{(\eps)}(\la,\eta ,k):=\left({\cal
    R}_{2}^{(\eps)}(\la,-\eta ,k)\right)^*.
$$
Let ${\frak w}_\eps^{(\iota)}(\la,\eta )$ be the column vectors obtained by scalar multiplication of each component
of ${\frak w}_\eps(\la,\eta ,k)$ by ${\frak e}_\iota$.
Note that
\begin{equation}
\label{061208}
\frak R_{\eps}=\frac{3\ga}{2}\sum_{\iota=\pm}{\frak e}_\iota F{\frak w}_\eps^{(-\iota)}(\la,\eta )+\eps^2{\frak h}_{\eps},
\end{equation}
where   the matrix
$F=(1/2){\rm e}^T\otimes {\rm e}$, vector ${\rm e}^T:=[1,-1,-1,1]$,
\begin{equation}
\label{h-eps}
{\frak h}_{\eps}(\la,\eta ,k)=\widehat{\frak W}_\eps (\eta
,k)-\frac{\ga (\pi \eta )^2}{2} f_\eps{\rm e}+{\frak r}_{\eps}(\la,\eta ,k)
\end{equation}
and $\widehat{\frak W}_\eps(\eta,k)$ is the column vector
corresponding to the Fourier-Wigner transforms of the components of   \eqref{032610},
and
\begin{equation}
\label{051208}
{\frak r}^T_\eps:=\left[r_{\eps}^{(1)},r_{\eps}^{(2)},r_{\eps,-}^{(2)},r_{\eps,-}^{(1)}\right],
 \end{equation}
$$
r_{\eps,-}^{(1)}(\la,\eta ,k):=r_{\eps}^{(1)}(\la,\eta ,-k), \qquad
r_{\eps,-}^{(2)}(\la,\eta ,k):=(r_{\eps}^{(2)}(\la,-\eta ,k))^*.
$$
Recall that ${\rm a}\otimes {\rm
  b}=[a_ib_j]$,
if ${\rm a}=[a_1,\ldots,a_n]$ and ${\rm b}=[b_1,\ldots,b_m]$.

\subsection{Invertibility of  matrix $\tilde D_\eps$}

We prove  that the matrix  $\tilde D_\eps$ appearing in
\eqref{011812}  is invertible, thus  the vector of the
Laplace-Fourier transforms of Wigner functions is uniquely determined
by the system. It turns out to be
true, provided that $\la$ is sufficiently large.

Let us denote
$\tilde\delta_\eps(\la,\eta,k):={\rm
  det}\,\tilde D_\eps(\la,\eta,k)$.
 Since matrices $B_\eps$ and $C_\eps$
commute we have (see p. 56 of \cite{gant})
\begin{eqnarray*}
&&\tilde\delta_\eps={\rm det}(A_\eps C_\eps-B^2_\eps)\\
&&
=\left|\tilde D_{1}^{(\eps)}\left(\tilde D_{2}^{(\eps)}\right)^*+[\tilde D_{+}^{(\eps)}]^2-[\tilde D_{-}^{(\eps)}]^2 \right|^2-4[\tilde D_{+}^{(\eps)}]^2{\rm Re}\,
\tilde D_{1}^{(\eps)}\,{\rm Re}\,
\tilde D_{2}^{(\eps)}.
\end{eqnarray*}

After a direct calculation we get
\begin{eqnarray}
\label{det-a}
&&
\tilde\delta_\eps
  =\eps^8\la^4 + 8\eps^6\la^3\ga R_\eps+ 4
\eps^4\la^2 \left[5(\ga R_\eps)^2+\bar\om^2+\left(\frac{\eps\delta_\eps\om}{2}\right)^2 -\left(\frac{\ga\eps R' \eta }{2}\right)^2 \right] \nonumber\\
&&+
 4\eps^2\la( \ga R_\eps)\left[ 4(\ga R_\eps)^2 + 4 \bar\om^2 +
   (\eps\delta_\eps\om)^2- \left(\ga\eps 
      R' \eta \right)^2  \right]
\\
&&
 + 4\eps^2\left[ (\ga R_\eps \delta_\eps\om )^2 - 2 (\ga R_\eps) \delta_\eps\om \bar\om \ga
      R' \eta + (\bar\om\delta_\eps\om )^2 \right]+ 16(\ga R_\eps \bar\om)^2.\nonumber
\end{eqnarray}

\bigskip

Define
\begin{equation}
\label{082312}
\tilde\delta_\eps^{(0)}:=(\eps^{2}\la + R_\eps )^4.
\end{equation}
\begin{prop}
\label{023112}
For any $M>0$ there exist $\eps_0(M),\la_0(M)>0$ such that 
\begin{equation}
\label{051612}
\tilde\delta_\eps(\la,\eta,k)\approx
\tilde\delta_\eps^{(0)}(\la,\eta,k), \,k\in\bbT,\,|\eta|\le M,\,\la>\la_0,\,\eps\in(0,\eps_0].
\end{equation}
In particular, we have
\begin{equation}
\label{051612a}
\tilde\delta_\eps(\la,\eta,k)>0, \,k\in\bbT,\,|\eta|\le M,\,\la>\la_0,\,\eps\in(0,\eps_0].
\end{equation}
\end{prop}
\proof 
Using \eqref{beta2} we conclude that
for any $M>0$ there is $\eps_0>0$ such that
\begin{equation}
\label{010407}
R_\eps \approx R(k)+(\eps \eta)^2,\quad k\in\bbT,\,|\eta|\le M,\,
\eps\in(0,\eps_0].
\end{equation}
Comparing the second formula from \eqref{050411} with \eqref{010407}
we get
\begin{equation}
\label{060307}
\bar \om\approx R_\eps ,\quad k\in\bbT,\,|\eta|\le M,\,
\eps\in(0,\eps_0].
\end{equation}
From \eqref{beta1}, the first formula of \eqref{050411} and
\eqref{060307} we get also
\begin{equation}
\label{051908}
|\delta_\eps\om R'|\preceq R_\eps,\quad |\eta|\le M, \, k\in\bbT.
\end{equation}
Therefore 
$$
8(\eps\ga)^2 R_\eps |\delta_\eps\om\bar \om
      R' \eta|\preceq \eps^2 R_\eps^3,\quad |\eta|\le M, \,k\in\bbT,\,\eps\in(0,\eps_0].
      $$
Choosing $\la_0$ sufficiently large we can guarantee also that  
\begin{equation}
\label{011908}
\eps^2 \la ( \ga R_\eps)^3\ge 8(\eps\ga)^2 R_\eps |\delta_\eps\om\bar \om
      R' \eta|
      \end{equation}
      for  $|\eta|\le M, \,k\in\bbT,\,\la>\la_0,\eps\in(0,\eps_0]$.

In a similar fashion we can argue that
\begin{equation}
\label{021908}
\eps^4\la^2  (\ga R_\eps)^2  \ge 4\eps^{4}\la( \ga R_\eps)\left[
   (\delta_\eps\om)^2- \left(\ga \hat
      R' \eta\right)^2  \right]
  \end{equation}
  and
  \begin{equation}
\label{031908}
\eps^6\la^3 \ga R_\eps\ge  
\eps^{6}\la^2 \left[\left(\delta_\eps\om\right)^2 -\left(\ga R' \eta\right)^2 \right] 
  \end{equation} 
      for   $|\eta|\le M, \,k\in\bbT,\,\la>\la_0,\eps\in(0,\eps_0]$.
From estimates  \eqref{011908}-\eqref{031908} we conclude 
that 
$$
\tilde\delta_\eps\succeq\eps^{8}\la^4 +
\eps^{6}\la ^3 R_\eps
+ 
\eps^4\la^2  R_\eps^2 +\eps^2\la R_\eps^3 + R_\eps ^4.
$$
Therefore, cf \eqref{082312}, we get $\tilde\delta_\eps^{(0)}\preceq\tilde\delta_\eps$. The reverse estimate is a simple consequence of
the first two formulas from \eqref{050411} and   \eqref{051908}.
 \qed

\bigskip


\bigskip

\subsection{Inverse of $\tilde D _\eps (\la,p,k)$}

Recall that $\tilde D _\eps (\la,p,k)$ is a $2\times 2$ block matrix
of the form  \eqref{011612}. Since $B_\eps$ is diagonal we
have $[A_\eps,B_\eps]=[C_\eps,B_\eps]=0$. A simple calculation shows that also
\begin{equation}
 \label{011612aae}
A_\eps C_\eps=C_\eps A_\eps.
  \end{equation}
Therefore,
\begin{equation}
 \label{011612aad}
 \tilde D^{-1}_\eps=
 \left[
 \begin{array}{cc}
(C _\eps  A _\eps-B _\eps^2)^{-1} &0\\
0&(C _\eps A-B _\eps^2)^{-1}
 \end{array}
 \right] \left[
 \begin{array}{cc}
C _\eps &-B _\eps \\
-B _\eps&A _\eps
 \end{array}
 \right],
 \end{equation}
provided that ${\rm det}\tilde D_\eps\not=0$. 
Note that
$$
(C_\eps A-B_\eps^2)^{-1}=\tilde\delta_\eps^{-1}\left[
 \begin{array}{cc}
\left(\tilde D_{1}^{(\eps)}\right)^*\tilde D_{2}^{(\eps)}+[\tilde D_{+}^{(\eps)}]^2-[\tilde D_{-}^{(\eps)}]^2  &-2\tilde D_{+}^{(\eps)}{\rm Re}\,
\tilde D_{2}^{(\eps)}\\
&\\
-2\tilde D_{+}^{(\eps)}{\rm Re}\, \tilde D_{1}^{(\eps)}&\tilde D_{1}^{(\eps)}\left(\tilde D_{2}^{(\eps)}\right)^*+[\tilde D_{+}^{(\eps)}]^2-[\tilde D_{-}^{(\eps)}]^2
 \end{array}
 \right].
$$
Substituting into \eqref{011612aad}, using also \eqref{012410}
we conclude that the inverse matrix $\tilde D^{-1}_\eps$ is a $2\times 2$ block matrix of the form
$\tilde D^{-1}_\eps=
\tilde\delta_\eps^{-1}{\rm adj}(\tilde D_\eps)$ where the adjugate of
$\tilde D_\eps$ equals
\begin{equation}
\label{011208}
{\rm adj}( \tilde  D_\eps)=
\left[\begin{array}{ll}
 P_\eps& Q_\eps\\
 Q_\eps&M_\eps
 \end{array}
 \right], 
\end{equation}
where $M_\eps$, $P_\eps$ and  $Q_\eps$ are $2\times 2$ matrices given by
\begin{eqnarray*}
 &&P_\eps:=\left[\begin{array}{ll}
 \tilde d_{1}^{(\eps)}&\tilde d_{-}^{(\eps)}\\
 \tilde d_{-}^{(\eps)}&\tilde d_{2}^{(\eps)}
 \end{array}
 \right],\quad
Q_\eps:=\left[\begin{array}{ll}
 (\tilde d_{+}^{(\eps)})^*&\tilde d_{o}^{(\eps)}\\
 \tilde  d_{o}^{(\eps)}&\tilde d_{+}^{(\eps)}
 \end{array}
 \right],\\
&&
M_\eps:=
\left[\begin{array}{ll}
(\tilde  d_{2}^{(\eps)})^*&(\tilde d_{-}^{(\eps)})^*\\
(\tilde d_{-}^{(\eps)})^*& (\tilde d_{1}^{(\eps)})^*
 \end{array}
 \right].
\end{eqnarray*}
Here
\begin{eqnarray}
\label{012910}
&&
\tilde d_1^{(\eps)}:=|\tilde D_2^{(\eps)}|^2 \left(\tilde D_1^{(\eps)}\right)^*-\left(\left(\tilde D_+^{(\eps)}\right)^2+ \left(\tilde
D_-^{(\eps)}\right)^2\right){\rm Re}\,\tilde D_2^{(\eps)}\nonumber\\
&&
-i \left(\left(\tilde D_+^{(\eps)}\right)^2- \left(\tilde
D_-^{(\eps)}\right)^2\right) {\rm Im}\,\tilde D_2^{(\eps)},\nonumber\\
&&
\tilde d_2^{(\eps)}:=\left|\tilde D_1^{(\eps)}\right|^2 \left(\tilde D_2^{(\eps)}\right)^*-\left(\left(\tilde D_+^{(\eps)}\right)^2+ \left(\tilde
D_-^{(\eps)}\right)^2\right){\rm Re}\,\tilde D_1^{(\eps)}\nonumber\\
&&
-i \left(\left(\tilde D_+^{(\eps)}\right)^2- \left(\tilde
D_-^{(\eps)}\right)^2\right) {\rm Im}\,\tilde D_1^{(\eps)},\\
&&
\tilde d_-^{(\eps)}:=\tilde D_+^{(\eps)} \left(\left(\tilde D_+^{(\eps)}\right)^2-\left( \tilde
D_-^{(\eps)}\right)^2\right)- \tilde D_+^{(\eps)}\left(\tilde D_1^{(\eps)}\tilde D_2^{(\eps)}\right)^*,\nonumber\\
&&
\tilde d_{+}^{(\eps)}:=-\tilde D_{-}^{(\eps)}\left(\tilde D_{1}^{(\eps)}\left(\tilde D_{2}^{(\eps)}\right)^*+\left(\tilde
D_{+}^{(\eps)}\right)^2-\left(\tilde D_{-}^{(\eps)}\right)^2\right),\nonumber\\
&&
\tilde d_{o}^{(\eps)}:= 2\tilde D_{+}^{(\eps)}\tilde D_{-}^{(\eps)}{\rm Re}\,
\tilde D_{2}^{(\eps)}. \nonumber
\end{eqnarray}
For the  abbreviation sake  we denote by ${\frak d}_{j,\eps}$, $j=1,\ldots,4$ the vectors
corresponding to the rows of the   adjugate of
$\tilde D_\eps$ given by \eqref{011208}.
Combining the above with \eqref{030911} and \eqref{det-a} we get.
\begin{prop}
\label{lm012008}
For any $M,\la>0$ we have
\begin{eqnarray}
\label{042008}
&&
\tilde d_{1}^{(\eps)}= 4(\ga R)^3 + 8\ga R \om^2 +o(1),\nonumber\\
&&
\tilde d_{2}^{(\eps)}= 4(\ga R)^3 - 8i(\ga
R)^2 \om +o(1),\nonumber\\
&&
\tilde d_{\pm}^{(\eps)}= 4(\ga R)^3 - 4i(\ga R)^2 \om +o(1),\\
&&
\tilde d_{o}^{(\eps)}= 4(\ga R)^3 +o(1),\nonumber\\
&&
\tilde\delta_\eps=16(\ga R\om)^2 +o(1),\quad \mbox{as }\eps\ll1, \nonumber
\end{eqnarray}
uniformly in $|\eta|\le M$ for any $k\in\bbT$.
\end{prop}

\section{Proof of Theorem \ref{main-thm}}

\label{sec7}

As we have already mentioned for any sequence $\eps_n\to0+$ there exists a subsequence $\left({\frak W}_{\eps_{n'}}(t)\right)$ that convergences 
 $*-$weakly to some 
 ${\frak W}\in\left( L^1([0,+\infty);{\cal A})\right)^*$. We prove that the element ${\frak W}$ does not depend on the choice of the sequence $\eps_{n'}$ by showing that for any $M>0$ there exists $\la_0>0$ such that the vector 
$\left({\frak w}_{\eps_{n'}}(\la)\right)$ made  of Laplace transforms 
of the components of ${\frak W}_{\eps_{n'}}(t)$ converges $*$-weakly over ${\cal A}_M'$ for any $\la>\la_0$. In fact one can describe the respective  limit 
as the Laplace transform of the vector ${\frak W}(t)$ appearing in the statement of Theorem \ref{main-thm}. This identifies the limit of   $\left({\frak W}_{\eps}(t)\right)$, as $\eps\to0+$ finishing in this way the proof of Theorem \ref{main-thm}.


From  \eqref{011812}  we obtain
\begin{equation}
\label{101110aa}
{\frak w}_\eps=\tilde D_\eps^{-1}\frak R_{\eps}.
\end{equation}
Unfortunately, the right hand side of the above system contains also
terms that depend on the vector ${\frak w}_\eps$, via the projections
of its components
onto the vectors ${\frak e}_{\pm}$ and ${\frak f}_{\pm}$. To describe
the behavior of
${\frak w}_\eps$ we need to determine first these projections. 

Using  \eqref{011812}  the above system can be rewritten in the form
\begin{equation}
\label{101110ab}
\frac{1}{\eps^2}\left({\frak w}_\eps-\frac{3\ga }{2
  }\sum_{\iota=\pm}\frak e_{\iota}\tilde E_\eps{\frak w}_\eps^{(-\iota)}\right)= {\frak z}_{\eps},
\end{equation}
where
\begin{equation}
\label{032610}
 {\frak z}_{\eps}^T(\la,\eta,k)=\left[
z_{\eps}^{(1)},
 z_{\eps}^{(2)},
 z_{\eps,-}^{(2)},
 z_{\eps,-}^{(1)}
 \right]:=\tilde D_\eps^{-1}\frak h_{\eps},
\end{equation}
  $\frak h_{\eps}$ is given by \eqref{051208} and the $4\times 4$
matrix $\tilde E_\eps(\la, \eta,k)$ equals 
\begin{eqnarray}
 \label{010201}
 \tilde E_\eps:=\frac12{\rm e}^T\otimes \Delta_\eps,
  \end{eqnarray}
with $\Delta_\eps^T:=[\tilde \Delta_{1,\eps},\tilde \Delta_{2,\eps}
,\tilde \Delta^*_{2,\eps},\tilde \Delta_{1,\eps}^*]$ and
\begin{eqnarray}
\label{D-1}
&&
\tilde \Delta_{1,\eps}:=\tilde d_1^{(\eps)}+\tilde d_{o}^{(\eps)}-\tilde d_{-}^{(\eps)}-(\tilde d_{+}^{(\eps)})^*,\nonumber\\
&&
\tilde \Delta_{2,\eps}:=\tilde d_{-}^{(\eps)}+\tilde d_{+}^{(\eps)}-\tilde d_2^{(\eps)}-\tilde d_{o}^{(\eps)}.
\end{eqnarray}
Multiplying both sides of \eqref{101110ab}  by $\frak e_{\iota}$, $\iota\in\{-,+\}$  and then integrating over $\bbT$ we get a system of $8$ equations
  \begin{eqnarray}
 \label{101110b}
 &&
 G_\eps{\frak u}_{\eps}=
 {\frak v}_{\eps},
  \end{eqnarray}
where
$$
{\frak u}_{\eps}(\la,\eta ):=\left[
 \begin{array}{l}
{\frak w}_{\eps}^{(-)}\\
 {\frak w}_{\eps}^{(+)}
  \end{array}
 \right],\qquad {\frak v}_{\eps}(\la,\eta ):=\left[
 \begin{array}{l}
{\frak z}_{\eps}^{(-)}\\
 {\frak z}_{\eps}^{(+)}
  \end{array}
 \right].
$$
Here
${\frak w}_{\eps}^{(\iota)}$
are column vectors obtained by a scalar multiplication of the entries
of  ${\frak w}_{\eps}$ (see \eqref{w}) by $\frak e_{\iota}$. The same
concerns 
\begin{equation}
\label{032610a}
\left( {\frak z}_{\eps}^{(\iota)}\right)^T(\la,\eta,k)=\left[
z_{\eps}^{(1,\iota)},
 z_{\eps}^{(2,\iota)},
 z_{\eps,-}^{(2,\iota)},
 z_{\eps,-}^{(1,\iota)}
 \right].
\end{equation}
Matrix $G_\eps(\la,\eta)$ is a $2\times 2$ block matrix of the form 
$$
G_\eps=\left[
 \begin{array}{cc}
A_{o}^{(\eps)}& A_{-}^{(\eps)} \\
&\\
A_{+}^{(\eps)}&A_{o}^{(\eps)} \end{array}
 \right],
$$
where $A_{o}^{(\eps)}$, $A_{\pm}^{(\eps)}$
are $4\times 4$ matrices defined as follows:
\begin{eqnarray*}
&& A_{\iota}^{(\eps)}:=-\frac{3\ga }{2
  \eps^{2}}\int_{\bbT}\frac{\frak e_{\iota}^2}{\tilde\delta_\eps}\tilde
E_\eps dk,\quad\iota\in\{-,+\},\\
&&
\\
&&
A_{o}^{(\eps)}:=\eps^{-2}\left(I-\frac{3\ga }{2
  }\int_{\bbT}\frac{\frak e_{-}\frak e_{+}}{\tilde\delta_\eps}\tilde E_\eps
dk\right).
\end{eqnarray*}
Note that vector ${\frak v}_{\eps}$ appearing on the right hand side of 
\eqref{101110b} still depends on the projections of ${\frak w}_\eps$
onto ${\frak f}_{\pm}$, cf \eqref{051208} and \eqref{f-eps}. It turns
out however that the asymptotics of these projections, as $\eps\to0+$, can be described
by only one of them, e.g. ${\frak w}_\eps^{(-)}$. This is a conclusion of
our next result.
Denote by  $\delta w_{\eps}:= w_{\eps}^{(+)}- w_{\eps}^{(-)}$. We
shall also use the following convention: for a given $M>0$ the
constants $\eps_0,\la_0>0$ are selected  as in the statement of Proposition
\ref{023112} so that $\tilde \delta_\eps(\la,\eta,k)\approx \tilde
\delta_\eps^{(0)}(\la,\eta,k)$ for all $k\in\bbT$, $|\eta|\le M$ and $\la>\la_0$.
In particular, then  we have \eqref{101110aa}.
\begin{thm}
\label{cor011811} 
For any $M>0$ and $\la>\la_0$ we have 
\begin{equation}
\label{011811}
|\delta w_{\eps}(\la,\eta )|\preceq \eps^{2}
\end{equation}
and 
 \begin{equation}
 \label{021811}
|  y_{\eps,\iota'}^{(\iota)}(\la,\eta )|\preceq
\eps^{2},\quad\,\eps\in(0,\eps_0],\,|\eta |\le M,\,\iota,\iota'\in\{-,+\}.
\end{equation}
Moreover,  for any $|\eta |\le M$ and $\la>\la_0$ we have
\begin{equation}
\label{031811}
\lim_{\eps\to0+}\int_{\bbT}\left| w_{\eps,+}(\la,\eta ,k)-
  w_{\eps}^{(-)}(\la,\eta )\right|R(k)dk
=0
\end{equation}
 and
\begin{equation}
\label{031811a}
\lim_{\eps\to0+}\int_{\bbT}\left| 
  y_{\eps,\pm}(\la,\eta ,k)\right|R(k)dk=0.
\end{equation}
\end{thm}
The proof of the theorem is presented in Section \ref{sec4}.

To describe the limit of $w_{\eps}^{(-)}(\la,\eta )$ we can use the
the system \eqref{101110b}, which is ''almost closed'' with respect to
the components of  ${\frak w}_\eps^{(-)}$, i.e. it is closed modulo
some corrections that in light of Theorem \ref{cor011811} are of lower
order of magnitude. 

Let us first introduce some additional notation. Given the wave
function $\phi(t,y)$ we define the vector of the Laplace-Fourier transforms of the respective macroscopic Wigner functions 
\begin{equation}
\label{032810}
{\frak w}^T_{\phi}(\la,\eta,h)=[{ w}_{\phi,+},{ y}_{\phi,+},{ y}_{\phi,-},{ w}_{\phi,-}],
\end{equation}
where 
\begin{eqnarray}
\label{022810}
&&
{ w}_{\phi,\pm}(\la,\eta, h):=\int_0^{+\infty}e^{-\la
  t}\widehat{ W}_{\phi,+}(t,\eta, h)dt,\\
&&
 { y}_{\phi,\pm}(\la,\eta, h):=\int_0^{+\infty}e^{-\la
  t}\widehat{ Y}_{\phi,+}(t,\eta, h) dt,\quad \la>0, \,(\eta,h)\in\bbR^2.\nonumber
\end{eqnarray}
Here 
\begin{eqnarray}
\label{011308}
&&
\widehat{ W}_{\phi,+}(t,\eta, h):=\frac12\hat \phi^*\left(t,h-\frac{ \eta}{2}\right) \hat
\phi\left(t,h+\frac{ \eta}{2}\right),\nonumber\\
&&
\\
&&
\widehat{ Y}_{\phi,+}(t,\eta, h):=\frac12\hat \phi\left(t,-h+\frac{ \eta}{2}\right) \hat
\phi\left(t,h+\frac{ \eta}{2}\right),\nonumber\\
&&
\widehat{ Y}_{\phi,-}
(t,\eta,k):=\left(\widehat{ Y}_{\phi,+}\right)^*(t,-\eta,h),\quad
\widehat{  W}_{\phi,-} (t,\eta,h):=\widehat{ W}_{\phi,+} (t,\eta,-h).\nonumber
\end{eqnarray}
Define  $\bar{\frak w}_{\phi}^T(\la,\eta):=\left[
 \overline w_{\phi},
\overline  y_{\phi,+},
\overline   y_{\phi,-},
 \overline  w_{\phi}
 \right],$
where
\begin{equation}
\label{wphi}
\overline  w_{\phi} (\la,\eta ):=\int_{\bbR}
{  w}_{\phi,\pm}(\la,\eta , h) dh,\quad \overline  y_{\phi,\pm} (\la,\eta ):=\int_{\bbR}
{ y}_{\phi,\pm}(\la,\eta , h) dh.
\end{equation}
Define $w^{(-)}(\la,\eta)$ by the formula
\begin{eqnarray}
\label{072008}
&&\left(\la+\hat c(2\pi \eta)^2\right)w^{(-)}(\la,\eta) =-3\ga(\pi\eta)^2\bar {\frak
w}_{\phi}(\la,\eta) \cdot {\rm e}\,\nonumber\\
&&
\\
&&
+6\ga\pi^2\int_{\bbR} h^2{\frak
w}_{\phi}(\la,\eta,h) \cdot {\rm e}\,dh+\hat e_{\rm th}(\eta), \quad
   (\la,\eta)\in(0,+\infty) \times\bbR.\nonumber
\end{eqnarray}
Here $\hat e_{\rm th}(\eta)$ is the Fourier transform of $e_{\rm
  th}(y)$ appearing in \eqref{072010a1} and $\hat c$ is given by
\eqref{hatc1}.

We can show, see Section \ref{sec11} below for the proof,
the following result.
\begin{thm}
\label{thm012410}
For any $M>0$ and $J\in {\cal S}(\bbR)$ such that ${\rm supp}\,\hat J\subset[-M,M]$ we have
$$
\int_{\bbR}w^{(-)}(\la,\eta )\hat J^*(\eta)d\eta=\lim_{\eps\to0+}\int_{\bbR}w_{\eps}^{(-)}(\la,\eta ) \hat J^*(\eta)d\eta
$$
for all $\la>\la_0$. 
\end{thm}

\bigskip

To obtain the asymptotics of ${\frak w}_{\eps}(\la,\eta,k )$ we use
\eqref{101110aa}, which allows us to describe the Fourier-Laplace
transforms of the Wigner functions in terms of their projections onto
${\frak e}_\pm$ and  ${\frak f}_\pm$. We obtain then the following result.
\begin{thm}
\label{thm012610}
 For any $M>0$  we have
\begin{equation}
\label{010707}
\lim_{\eps\to0+}\left|\int_{\bbT}
  w_{\eps,+}(\la,\eta ,k)\varphi(k)dk-w^{(-)}_\eps(\la,\eta )\int_\bbT\varphi(k)dk-
   \overline w_{\phi} (\la,\eta )\varphi(0)\right|=0
\end{equation}
and
\begin{equation}
\label{031811b}
\lim_{\eps\to0+}\int_{\bbT}
  y_{\eps,\pm}(\la,\eta ,k)\varphi(k)dk=
   \overline  y_{\phi,\pm} (\la,\eta )\varphi(0),\quad
\end{equation}
 for all $|\eta |\le M$, $\la>\la_0$ and $\varphi\in C(\bbT)$.
\end{thm}
The proof of this result is contained in Section \ref{sec4a}.

\bigskip

\subsection*{The end of the proof of Theorem \ref{main-thm}}

Thanks to \eqref{W-Y} we know that ${\frak W}_\eps(t)$ is sequentially pre-compact , as $\eps\to0+$, in the $*$-weak topology
 of $\left(L^1([0,+\infty),{\cal A})\right)^*$. To identify its limiting points we consider ${\frak w}_{\eps}(\la,\eta,k )$ 
 the vector of the Laplace-Fourier transforms of ${\frak W}_\eps(t)$. Given  $\la>0$
 this family is sequentially pre-compact in  the $*$-weak topology opf ${\cal A}'$, as $\eps\to0+$.
Thanks to Theorems \ref{thm012410} and  \ref{thm012610}
we conclude that given $M>0$ one can choose $\la_0$ as in the
statement of Proposition \ref{023112},   such that  the the components of
${\frak w}_{\eps}(\la,\eta,k )$ 
converge $*$-weakly over ${\cal A}'_M$ to the Laplace-Fourier transforms of the
respective functions appearing in the claim of Theorem \ref{main-thm} for any $\la>\la_0$.
To finish the proof we only need to verify that  $ w(\la,d\eta,dk)$ - the limit of  $
w_{\eps,+}(\la,\eta,k)$ (the limit of $
w_{\eps,-}$ can then be trivially concluded) agrees for $\la>\la_0$
with the Laplace
transform of $W(t,dy,dk)$ appearing in \eqref{032910}.

According to Theorem \ref{thm012610} the limit in question is the Fourier-Laplace
transform  of the measure-valued function
$$
W'(t,dy,dk)= e'_{\rm th}(t,y)dydk+e_{\rm mech}(t,y)dy\delta_0(dk),
$$
where, according to \eqref{072008}, we have
\begin{align}
\label{012610}
&e'_{\rm th}(0,y)=e_{\rm th}(y),\\
&
\partial_te'_{\rm th}(t,y)=\hat c\partial_y^2e'_{\rm th}(t,y)+\frac{3\ga}{4}\partial_y^2\bar {\frak w}_{\phi}(t,y) \cdot {\rm e}+12\ga
\pi^2\int_{\bbR} h^2  {\frak W}_{\phi}(t,y,h) \cdot {\rm e}\,dh.\nonumber
\end{align}
Here $\bar {\frak w}_{\phi}(t,y)$ is defined in \eqref{wphi}, 
$$
 {\frak W}_{\phi}^T(t,y,k):=[ { W}_{\phi,+}(t,y,k), { Y}_{\phi,+}(t,y,k),{ Y}_{\phi,-}(t,y,k),{ W}_{\phi,-}(t,y,k)]
$$
and
\begin{equation*}
{ W}_{\phi,\pm}(t,y,k)=\int_{\bbR}e^{2\pi i y \eta} \widehat{
  W}_{\phi,\pm}(t,\eta,k)d\eta,\quad  { Y}_{\phi,\pm}(t,y,k)=\int_{\bbR}e^{2\pi i y \eta} \widehat{
  Y}_{\phi,\pm}(t,\eta,k)d\eta,
\end{equation*}
with $\widehat{
  W}_{\phi,\pm}$ and $\widehat{
  Y}_{\phi,\pm}$  given by \eqref{011308}.
An elementary calculation yields the following.
\begin{prop}
\label{prop011308}
Suppose that $\phi(t,y)$ is given by \eqref{012810}. Then, 
\begin{eqnarray*}
&&
\int_{\bbR} {W}_{\phi,\pm}(t,y,h) dh = \frac12|\phi(t,y)|^2,\\
&&
\int_{\bbR} {Y}_{\phi,+}(t,y,h) dh = \frac{1}{2}[\phi(t,y)]^2,\\
&&
4\pi^2\int_{\bbR} h^2{W}_{\phi,\pm}(t,y,h) dh = 
\frac{1}{4}|\phi'(t,y)|^2-\frac18\left[\phi(t,y)(\phi'')^*(t,y)+\phi''(t,y)\phi^*(t,y)\right],\\
&&
4\pi^2\int_{\bbR} h^2 {Y}_{\phi,+}(t,y,h) dh = \frac{1}{4}\left\{[\phi'(t,y)]^2-\phi(t,y)\phi''(t,y)\right\}
\end{eqnarray*}
 for any $(t,y)\in [0,+\infty)\times\bbR$.
\end{prop}
Using the proposition we conclude that the third term appearing in
the right hand side of the second equation of \eqref{012610}
equals
\begin{eqnarray*}
&&
12\ga\pi^2
\int_{\bbR} h^2 \left( { W}_{\phi,+}(t,y,h) - {\rm Re}\, {
    Y}_{\phi,+}(t,y,h) \right) dh\\
&&
=\frac{3\ga}{2}
\left[(p')^2(t,y)-p(t,y)p''(t,y)\right].
\end{eqnarray*}
On the other hand, the second term equals
\begin{eqnarray*}
&&
\frac{3\ga}{4}\partial_y^2\left\{\int_{\bbR}  \left( { W}_{\phi,+}(t,y,h) - {\rm Re}\, {
    Y}_{\phi,+}(t,y,h) \right)
dh\right\}=\frac{3\ga}{4}\partial_y^2p^2(t,y)\\
&&
=\frac{3\ga}{2}\left\{(p')^2(t,y)+p(t,y)p''(t,y)\right\}.
\end{eqnarray*}
We can see therefore that $e'_{\rm th}(t,y)$ satisfies
\eqref{energy-eqt}. Thus the conclusion of Theorem \ref{main-thm} follows.

\section{Proof of Theorem \ref{cor011811}}

\label{sec4}

We start with the following result.
\begin{lm}
\label{lm011912}
For any $M>0$ and $\eps_0,\la_0$ as in Proposition \ref{023112}
we have
\begin{equation}
\label{080912}
\frac{
  R}{\tilde\delta_\eps}\sum_j|\tilde d_j^{(\eps)}|\preceq
1,\quad\forall\,k\in\bbT,\,|\eta|\le M,\,\eps\in(0,\eps_0],\,\la>\la_0.
\end{equation}
The summation extends over $j\in\{1,2,o,-,+\}$.
\end{lm}
\proof 
From the definition of $\tilde D_{1}^{(\eps)}$, see
\eqref{030911}, we obtain
\begin{equation}
\label{022911ax1a}
|\tilde D_{1}^{(\eps)}|\preceq  R_\eps +   
\la\eps^{2}+\eps|\delta_\eps\om|
\end{equation}
for $k\in\bbT,\,|\eta|\le M,\,\eps\in(0,\eps_0],\la>\la_0$.
Using the first formula of \eqref{050411} we conclude that then
\begin{equation}
\label{022911ax1b}
|\tilde D_{1}^{(\eps)}|\preceq  R_\eps +   
\la\eps^{2}.
\end{equation}
A similar consideration leads to the estimate
\begin{equation}
\label{022911ax1}
|\tilde D_{j}^{(\eps)}|\preceq  R_\eps +   
\la\eps^{2}, \quad \,k\in\bbT,\,|\eta|\le M,\,\eps\in(0,\eps_0],\la>\la_0
\end{equation}
for any $j\in\{1,2, +,-\}$.
In particular we can conclude from \eqref{012910} that
\begin{equation}
\label{012911aax1}
\sum_j|\tilde d_j^{(\eps)}| \preceq (R_\eps +   
\la\eps^{2})^3 \approx \sum_{j=0}^3R_\eps^{3-j}(\la\eps^{2})^j .
\end{equation}
Thanks to  \eqref{051612}  we infer that
$$
R|\tilde d_{j}^{(\eps)}|\preceq (R_\eps +   
\la\eps^{2})^4\preceq \tilde\delta_\eps,\quad j\in\{1,2,o,+,-\}.
$$
\qed

\bigskip

\subsection{Proof of (\ref{021811})}

We show \eqref{021811} for $(\iota,\iota')=(-,+)$.
The cases of
 other values of $(\iota,\iota')$ can be
 handled in the same way.
We use the second
equation of the system \eqref{101110b}.
Estimate in question follows, provided we can show that the left hand
side of the equation
can be written in the form 
\begin{equation}
\label{032011}
\eps^{-2}   y_{\eps}^{(-)}(\la,\eta)+\tilde T_\eps(\la,\eta)=z_{\eps}^{(2,-)}(\la,\eta),
\end{equation}
where 
\begin{eqnarray}
\label{012111}
&&
\tilde T_\eps(\la,\eta)=O(1), \\
&&
z_{\eps}^{(2,-)}(\la,\eta)=O(1),\quad \mbox{ as }\eps\ll1\nonumber. 
\end{eqnarray}


\label{sec4.3}
We can write
\begin{equation}
\label{011911}
\tilde T_\eps(\la,\eta)=-2b_o^{(\eps)}v_{\eps}^{(-)}-2b_-^{(\eps)}v_{\eps}^{(+)},
\end{equation}
where
$
v_{\eps}^{(\pm)}(\la,\eta):=\langle v_{\eps}(\la,\eta,\cdot),{\frak
  e}_{\pm}\rangle
$  (see \eqref{v})
and
\begin{eqnarray}
\label{b}
&&
b_o^{(\eps)}:=-\frac{3\ga }{4 \eps^{2}}\int_{\bbT}\frak
e_{-} \frak
e_{+}\frac{\tilde \Delta_{2,\eps}
  dk}{\tilde\delta_\eps},\\
&&
b_\pm^{(\eps)}:=-\frac{3\ga }{4 \eps^{2}}\int_{\bbT}\frak
e_{\pm}^2\frac{\tilde \Delta_{2,\eps}dk}{\tilde\delta_\eps}.\nonumber
\end{eqnarray}
Substituting from \eqref{030911} into \eqref{D-1} we find
\begin{eqnarray}
\label{D-2a}
&&
\eps^{-2}\tilde \Delta_{2,\eps}
= \delta_\eps\om\left[ 2\eta \bar\om R' -
2(\ga R_\eps)\delta_\eps\om \right]\nonumber\\
&&
-4\la (\ga R_\eps)\left[ (\ga R_\eps) + \eps^{2}\la\right] +\eps^{2}\la\left[(\delta_\eps\om)^2- 
\left(R'\eta\right)^2\right]+\eps^{4}\la^3\nonumber\\
&&
\\
&&
+i\left\{\vphantom{\int_0^1}4\la \bar \om (\ga R_\eps)
+\eps^2\la \delta_\eps \om R'\eta+2\eps^2\bar\om(\delta_\eps\om)^2 + 2\eps^{2}\la ^2\bar\om\right\}.\nonumber
\end{eqnarray}
Therefore, cf \eqref{082312}, we conclude that
\begin{equation}
\label{D-2b}
\eps^{-2}|\tilde \Delta_{2,\eps}|\left({\frak e}_{-} \frak
e_{+}+{\frak e}_{-}^2 +{\frak e}_{+}^2\right)\preceq \eps^{-2}|\tilde \Delta_{2,\eps}|R^2
\preceq \tilde\delta_\eps^{(0)}.
\end{equation}
Thus,
\begin{equation}
\label{b-ep}
|b_o^{(\eps)}|+|b_-^{(\eps)}|+|b_+^{(\eps)}|\preceq 1
\end{equation}
and the first equality of \eqref{012111} follows.



Since (see \eqref{f-eps})
$$
f_{\eps}^*(\la,-\eta,k)=f_{\eps}(\la,\eta,-k) =f_{\eps}(\la,\eta,k)
$$ 
the right hand side of the second equation of system \eqref{101110b} can be written as
\begin{equation}
\label{z2}
 z_{\eps}^{(2,-)}=Z_{\eps,1}+Z_{\eps,2}+Z_{\eps,3},
\end{equation}
where     (${\frak d}_{j,\eps}$ are the 
rows of the adjugate matrix to $\tilde D_\eps$, given by  \eqref{011208})
\begin{eqnarray}
\label{012311}
&&
Z_{\eps,1}:=\int_{\bbT}
{\frak d}_{2,\eps}\cdot \widehat{\frak W}_\eps \frac{\frak e_{-}dk}{\tilde\delta_\eps} ,\nonumber\\
&&
\\
&&
Z_{\eps,2}:=-\frac{\ga (\pi \eta)^2}{2}\int_{\bbT}\tilde \Delta_{2,\eps}
\frac{\frak e_{-}
 f_{\eps} }{\tilde \delta_\eps}dk
, \nonumber\\
&&
\nonumber\\
&&
Z_{\eps,3}:=\eps\int_{\bbT}
{\frak d}_{2,\eps}\cdot{\frak r}_\eps  \frac{\frak e_{-}dk}{\tilde\delta_\eps}. \nonumber
\end{eqnarray}
We can write
\begin{equation}
\label{012008}
|Z_{\eps,1}|\le \left(\sum_j\sup_{k}\frac{\frak e_{-}|\tilde d_{j}^{(\eps)}|}{|\tilde\delta_\eps|}
\right)\sum_{\iota=\pm}\left(\|\widehat
W_{\eps,\iota}\|_{{\cal A}'} +
\|\widehat
Y_{\eps,\iota}\|_{{\cal A}'} \right). 
\end{equation}
Using Lemma \ref{lm011912} we conclude that for any $M>0$ there exists
$\la_0$ such that for any $\la>\la_0$ we have  
$|Z_{\eps,1}|=O(1)$, as $\eps\ll1$. A similar argument allows us to conclude that
also $|Z_{\eps,j}|=O(1)$, as $\eps\ll1$ for $j=2,3$.
Thus,
 the second equality in \eqref{012111} follows as well.

\bigskip

\subsection{Proof of (\ref{011811})}

The left hand side of the first equation of the system
\eqref{101110b} 
can be rewritten in the following form
\begin{equation}
\label{042811}
a^{(\eps)}_{w,-} w_{\eps}^{(-)}-a^{(\eps)}_-\frac{\delta w_{\eps}}{\eps^{2}}+a^{(\eps)}_o\frac{ y_{\eps,o}^{(-)}}{\eps^{2}}+a^{(\eps)}_-\frac{ y_{\eps,o}^{(+)}}{\eps^{2}}
\end{equation}
with
\begin{eqnarray}
\label{040201}
&&
a^{(\eps)}_{w,\pm}:=\eps^{-2}\left[1-2\int_{\bbT}(\ga R)\frak e_\pm
  \frac{\tilde \Delta_{1,\eps}}{\tilde\delta_\eps}dk\right], \\
&&
a^{(\eps)}_o:=\frac{3\ga}{4}\int_{\bbT}\frak e_{-}\frak e_+
  \frac{\tilde \Delta_{1,\eps}}{\tilde\delta_\eps}dk,\nonumber\\
&&
a^{(\eps)}_\pm:=\frac{3\ga}{4}\int_{\bbT}\frak e_{\pm}^2
  \frac{\tilde \Delta_{1,\eps}}{\tilde\delta_\eps}dk.\nonumber
\end{eqnarray}
Note that
$
\frak e_{-}\frak e_+\preceq R^3
$ (see \eqref{frak-e-r}).
From Lemma \ref{lm011912} and the Lebesgue
 dominated convergence theorem we conclude that $a^{(\eps)}_o$ and
$a^{(\eps)}_\pm$ are of order $O(1)$, as
$\eps\ll1$.

Using \eqref{D-1} together with formulas \eqref{042008}  we infer that
\begin{eqnarray}
\label{D-1e}
&&
\tilde \Delta_{1,\eps}=8\ga R_\eps\bar\om^2+o(1) ,\quad \mbox{as }\eps\ll1. 
\end{eqnarray}
Therefore, by the Lebesgue dominated convergence theorem  
$$
\lim_{\eps\to0+}a^{(\eps)}_o=\bar a_o:=\frac{3}{8}\int_{\bbT}\frac{\frak e_{-}\frak e_+
  dk}{R}>0
$$
and
$$
\lim_{\eps\to0+}a^{(\eps)}_\pm=\bar a_\pm:=\frac{3}{8}\int_{\bbT}\frac{\frak e_{\pm}^2
  dk}{R}>0.
$$
After a direct computation we obtain
$$
a^{(\eps)}_{w,-}=\int_{\bbT}\frak e_-
  \frac{\tilde e_{\eps}}{\tilde\delta_\eps}dk,
$$
where
\begin{eqnarray*}
&&
\tilde e_{\eps}:=\left\{4(\ga R)\eta  \delta_\eps\om\bar\om
R' + 2\eta ^2\ga
      R''(\ga R_\eps) \bar\om^2 + 4\left[ (\ga R_\eps \delta_\eps\om )^2 - 2 (\ga R_\eps) \delta_\eps\om \bar\om \hat
      R' \eta + (\bar\om\delta_\eps\om )^2\right]\right\}
\\
&&
+
 4\la( \ga R_\eps)\left[ 4(\ga R_\eps)^2 + 4 \bar\om^2 +
   (\eps\delta_\eps\om)^2- \left(\eps \hat
      R' \eta \right)^2  \right]
\\
&&
-4(\ga R)\la\left[2 (\ga R_\eps)^2
+ 2\bar\om^2-2\left(\frac{\eps R' \eta }{2}\right)^2\right]
\nonumber\\
&&
+ 4 \eps^{2} \left\{
\la^2 \left[5(\ga R_\eps)^2+\bar\om^2+(\eps\delta_\eps\om)^2 -\left(\frac{\eps \hat
      R' \eta }{2}\right)^2 \right] -2\ga R\la^2(\ga R_\eps)\right\}
\\
&&
-2(\ga R) \la^3\eps^{4}+ 8\eps^{4}\la ^3\ga R_\eps +\eps^{6}\la^4.\nonumber
\end{eqnarray*}
Taking into account \eqref{051612} we conclude 
that
$
\frak e_-\tilde e_{\eps}\preceq \tilde\delta_\eps.
$
In addition,
\begin{eqnarray*}
&&
\tilde e_{\eps}= 2\eta ^2\ga
      R''(\ga R_\eps) \bar\om^2      
+
8\la (\ga R_\eps)[(\ga R_\eps)^2 + \bar\om^2 ]
+o(1),\quad \mbox{as }\eps\ll1.
\end{eqnarray*}
Combining this with the second formula of \eqref{D-1e} we obtain,
by the Lebesgue
dominated convergence theorem, 
\begin{equation}
\label{032811a}
\bar a:=\lim_{\eps\to0+}\ga a^{(\eps)}_{w,-}<+\infty.
\end{equation}

Using the above together with bound \eqref{021811} 
 we conclude  that
expression \eqref{042811} can be written as 
\begin{equation}
\label{042811a}
\frac{\bar a}{\ga}  [1+o(1)] w_{\eps}^{(-)}-\bar a_-  [1+o(1)]\frac{\delta w_{\eps}}{\eps^{2}}+O(1) ,\quad \mbox{as }\eps\ll1.
\end{equation}
Then   bound \eqref{011811} would follow, provided we can show that  the right hand
side of the first equation of the system \eqref{101110b}, given by
$
z_{\eps}^{(1,-)}$, is of order of magnitude $O(1)$, as $\eps\ll1$.
To see that we write
\begin{equation}
\label{z1-a}
 z_{\eps}^{(1,-)}=U_{\eps,1}+U_{\eps,2}+U_{\eps,3},
\end{equation}
where the terms $U_{\eps,i}$, $i=1,2,3$
are given by
\begin{eqnarray}
\label{012311a}
&&
U_{\eps,1}=\int_{\bbT}
{\frak d}_{1,\eps}\cdot \widehat{\frak W}_{\eps} 
\frac{\frak e_{-} dk}{\tilde\delta_\eps} ,\nonumber\\
&&
\\
&&
U_{\eps,2}:=-\frac{\ga (\pi \eta)^2}{2}\int_{\bbT}\tilde \Delta_{1,\eps}
\frac{\frak e_{-}
 f_{\eps}}{\tilde \delta_\eps} dk
, \nonumber\\
&&
U_{\eps,3}:=\eps
\int_{\bbT}
{\frak d}_{1,\eps}\cdot{\frak r}_{\eps} 
\frac{\frak e_{-} dk}{\tilde\delta_\eps} 
. \nonumber
\end{eqnarray}
The fact that $z_{\eps}^{(1,-)}=O(1)$, as $\eps\ll1$, can be
argued in a similar way as it has been done in the case of $
z_{\eps}^{(2,-)}$, see \eqref{012311} and \eqref{012008} above.

\subsection{Proof of (\ref{031811})}

From \eqref{011812} we obtain
$$
  w_{\eps,+}(\la,\eta,k)=I_\eps+I\!I_\eps+I\!I\!I_\eps+I\!V_\eps,
$$
where
\begin{eqnarray}
\label{w-eps}
&&
I_\eps:=\frac{3\ga}{2\tilde \delta_{\eps}}\tilde \Delta_1^{(\eps)} \sum_{\iota\in\{-,+\}}\langle v_\eps,\frak e_{\iota}\rangle_{L^2(\bbT)} \frak e_{-\iota},\\
&&
\nonumber\\
&&
I\!I_\eps:=-\frac{\ga (\pi\eps \eta)^2}{2\tilde \delta_{\eps}}\tilde
\Delta_1^{(\eps)}\left[\frak f_+\langle v_\eps,16\frak f_++\frak
  e_{-}\rangle_{L^2(\bbT)}+\frak e_-\langle v_\eps,\frak
  f_{+}\rangle_{L^2(\bbT)}\right. \nonumber\\
&&
\left.+
3\frak f_{-}\langle v_\eps,\frak e_{+}\rangle_{L^2(\bbT)}+
3\frak e_{+}\langle v_\eps,\frak f_{-}\rangle_{L^2(\bbT)}\right],\\
&&
I\!I\!I_\eps:= \eps^{2}\tilde \delta_{\eps}^{-1}{\frak d}_{\eps,1}\cdot\widehat{\frak W}_{\eps},
\qquad
I\!V_\eps:=\eps^{3}\tilde \delta_{\eps}^{-1}{\frak d}_{\eps,1}\cdot {\frak r}_{\eps}.\nonumber
\end{eqnarray}

\subsubsection{Convergence of $I_\eps$}

\label{sec15.3.1}

Note that, (see \eqref{det-a}) for any $k\not=0$ 
\begin{equation}
\label{042910}
\lim_{\eps\to+0}\frac{(\ga R) \tilde \Delta_1^{(\eps)}}{\tilde\delta_{\eps}}=\frac12.
\end{equation}
Using the above and the already proved estimates \eqref{011811},   \eqref{021811} and
Lemma \ref{lm011912}
 we obtain 
\begin{equation}
\label{021112}
\lim_{\eps\to0+}\left\|I_\eps-w_{\eps}^{(-)}\frak f\right\|_{L^1(\bbT)}=0.
\end{equation}
Here $\frak f(k)\equiv1$.

\subsubsection{Convergence of $I\!I_\eps$, $I\!I\!I_\eps$ and $I\!V_\eps$}

\label{sec15.3.2}

Thanks to Lemma \ref{lm011912} we can write
$$
\eps^2\int_{\bbT}|\tilde \delta_{\eps}^{-1}\widehat W_\eps\tilde
d_1^{(\eps)}|R dk\preceq \eps^2\|\widehat W_\eps\|_{{\cal A}'}\to 0,\quad \mbox{as }\eps\to0+.
$$
The remaining terms appearing in expressions $I\!I_\eps$,
$I\!I\!I_\eps$ and $I\!V_\eps$
can be estimated in the same manner allowing us to conclude that
$$
\lim_{\eps\to0+}\int_{\bbT}\left(|I\!I_\eps|+|I\!I\!I_\eps|+|I\!V_\eps|\right)Rdk=0.
$$

\subsection{ Proof of $(\ref{031811a})$}

\label{sec8.5}

From \eqref{011812} we obtain
$$
 y_{\eps}(\la,\eta,k)=I_\eps+I\!I_\eps+I\!I\!I_\eps+I\!V_\eps,
$$
where
\begin{eqnarray*}
&&
I_\eps:=\frac{3\ga}{2\tilde \delta_{\eps}}\tilde \Delta_2^{(\eps)} \sum_{\iota\in\{-,+\}}\langle v_\eps,\frak e_{\iota}\rangle_{L^2(\bbT)} \frak e_{-\iota},\\
&&
 I\!I_\eps:=-\frac{\ga (\pi\eps \eta)^2}{2\tilde \delta_{\eps}}\tilde \Delta_2^{(\eps)}\left[\frak f_+\langle v_\eps,16\frak f_++\frak
  e_{-}\rangle_{L^2(\bbT)}+\frak e_-\langle v_\eps,\frak
  f_{+}\rangle_{L^2(\bbT)}\right. \nonumber\\
&&
\left.+
3\frak f_{-}\langle v_\eps,\frak e_{+}\rangle_{L^2(\bbT)}+
3\frak e_{+}\langle v_\eps,\frak f_{-}\rangle_{L^2(\bbT)}\right],\\
&&
I\!I\!I_\eps:= \eps^{2}\tilde \delta_{\eps}^{-1}{\frak d}_{\eps,2}\cdot\widehat{\frak W}_{\eps},
\qquad
I\!V_\eps:=\eps^{3}\tilde \delta_{\eps}^{-1}{\frak d}_{\eps,2}\cdot {\frak r}_{\eps}.\nonumber
\end{eqnarray*}
The analysis of the above terms is very similar to what has been done
in the precious section.
Using \eqref{D-2a} we conclude that for any $\la>\la_0$ 
\begin{equation}
\label{052910}
\left|\frac{\tilde \Delta_2^{(\eps)}(\la,\eta,k)}{\tilde
      \delta_{\eps}(\la,\eta,k)}\right|\preceq \eps^2,\quad k\in\bbT,\,|\eta|\le M.
\end{equation}   
We conclude in this way that all
$RI_\eps$, $R I\!I_\eps$, $R I\!I\!I_\eps$ and $RI\!V_\eps$   tend to $0$ in the $L^1$ sense. Thus,
\eqref{031811a} follows.

\bigskip

\section{Proof of Theorem \ref{thm012410}}

\label{sec11}

\subsection{Determining  $w_{\eps}^{(-)}$}

Since  
 functions $\frak e_{\pm}(k)$ are both even
the fourth and eighth equation of 
 the system  \eqref{101110b} coincide with the first and the fifth
 ones respectively.

Adding the first and fifth equations of the system \eqref{101110b}  we get
\begin{equation}
\label{020512}
\ga a^{(\eps)}_{w}  w_{\eps}^{(-)}-\sum_{\iota\in\{-,+\}}a^{(\eps)}_{y,-\iota}\hat
y_{\eps,o}^{(\iota)}= \ga  z_{\eps}^{(1,o)}+\ga a^{(\eps)}_{w,-} \delta w_{\eps}
\end{equation}
Here $a^{(\eps)}_{w,\iota}$, $a^{(\eps)}_{\iota}$, $\iota\in\{o,-,+\}$ are given
by \eqref{040201} and 
$
\hat y_{\eps,o}^{(\pm)}:=\eps^{-2}  y_{\eps,o}^{(\pm)}.
$
In addition
\begin{eqnarray*}
&&
a^{(\eps)}_{w}:=a^{(\eps)}_{w,-}+a^{(\eps)}_{w,+}=\frac{4}{3\eps^{2}}\int_{\bbT}\ga
R\left[1-2(\ga
  R)\frac{\tilde \Delta_{1,\eps}}{\tilde \delta_\eps}\right]
dk,\\
&&
a^{(\eps)}_{y,\pm}
:=a^{(\eps)}_{\pm}+a^{(\eps)}_o:=\int_{\bbT}{\frak e}_\pm\tilde \Delta_{1,\eps}\frac{\ga
  R  dk}{\tilde \delta_\eps}
\end{eqnarray*}
and
\begin{equation}
\label{z1}
 z^{(1,o)}_\eps:= z_{\eps}^{(1,-)}+z_{\eps}^{(1,+)},
 \end{equation}
where $z_{\eps}^{(1,\pm)}$ are the scalar products of $z_{\eps}^{(1)}$ by ${\frak e}_\pm$ (cf \eqref{032610}
and    \eqref{032610a}).

The second  and third equations of \eqref{101110b} read (cf \eqref{b})
\begin{eqnarray}
\label{010512}
&&
 \ga b_{o}^{(\eps)} w_{\eps}^{(-)}+\ga
b_-^{(\eps)} w_{\eps}^{(+)}+\hat y_{\eps,+}^{(-)} = \ga z_{\eps,+}^{(2,-)}-\eps^{2}\left[b_o^{(\eps)}\hat
  y_{\eps,o}^{(-)}+b_-^{(\eps)}\hat y_{\eps,o}^{(+)}\right],\nonumber
\\
&&
\\
&&
\ga (b_{o}^{(\eps)})^* w_{\eps}^{(-)}+\ga
(b_-^{(\eps)} )^*w_{\eps}^{(+)}+\hat y_{\eps,+}^{(-)} = \ga z_{\eps,-}^{(2,-)}-\eps^{2}\left[(b_o^{(\eps)})^*\hat
  y_{\eps,o}^{(-)}+(b_-^{(\eps)})^*\hat y_{\eps,o}^{(+)}\right].\nonumber
\end{eqnarray}
 Adding sideways these equations  we get
 \begin{equation}
\label{060512}
2\ga w_\eps^{(-)} {\rm Re}\,b_{o}^{(\eps)}+2\ga w_\eps^{(+)}{\rm Re}\,b_{-}^{(\eps)}+\hat y_{\eps,o}^{(-)}=\ga 
z_{\eps,o}^{(2,-)} +r_\eps^{(-)}.
\end{equation}
Here
$z_{\eps,o}^{(2,\pm)}:=z_{\eps}^{(2,\pm)}+ z_{\eps,-}^{(2,\pm)} $ 
and
\begin{equation}
\label{r-}
r_\eps^{(-)}:=-2\eps^{2}\left(\hat
  y_{\eps,o}^{(-)}{\rm Re} \,b_o^{(\eps)}+\hat y_{\eps,o}^{(+)}{\rm Re}  \,b_-^{(\eps)}\right),
 \end{equation}
 The sixth and seventh equations of \eqref{101110b} yield
 \begin{equation}
\label{010612}
2\ga w_\eps^{(-)}{\rm Re}\,b_{+}^{(\eps)}+2\ga w_\eps^{(+)}{\rm Re}\,b_{o}^{(\eps)}+\hat y_{\eps,o}^{(+)}=\ga 
z_{\eps,o}^{(2,+)} +r_\eps^{(+)},
\end{equation}
and 
\begin{equation}
\label{r+}
r_\eps^{(+)}:=-2\eps^{2}\left(\hat
  y_{\eps,o}^{(-)}{\rm Re} \,b_+^{(\eps)}+\hat y_{\eps,o}^{(+)}{\rm Re}  \,b_o^{(\eps)}\right).
 \end{equation}
Summarizing, we have obtained the following system 
\begin{eqnarray}
\label{sys-1}
&&
\ga a^{(\eps)}_{w}  w_{\eps}^{(-)}-\sum_{\iota\in\{-,+\}}a^{(\eps)}_{y,-\iota}\hat
y_{\eps,o}^{(\iota)}= \ga  z_{\eps}^{(1,o)}+\ga a^{(\eps)}_{w,-} \delta w_{\eps},\nonumber
\\
&&
2\ga w_\eps^{(-)}{\rm Re}\,b_{o}^{(\eps)}+2\ga w_\eps^{(+)}{\rm Re}\,b_{-}^{(\eps)}+\hat y_{\eps,o}^{(-)}=\ga 
z_{\eps,o}^{(2,-)} +r_\eps^{(-)},
\\
&&
\nonumber\\
&&
2\ga w_\eps^{(-)}{\rm Re}\,b_{+}^{(\eps)}+2\ga w_\eps^{(+)} {\rm Re}\,b_{o}^{(\eps)}+\hat y_{\eps,o}^{(+)}=\ga 
z_{\eps,o}^{(2,+)} +r_\eps^{(+)}.
\nonumber
\end{eqnarray}

Using Theorem \ref{cor011811} 
 we conclude that given $M>0$ and $\la>\la_0$ the family
$
( w_{\eps}^{(-)},\hat y^{(-)}_{\eps,o},\hat
y^{(+)}_{\eps,o})
$
remains bounded in $L^\infty[-M,M]$, as $\eps\to0+$. It is therefore $*$-weakly sequentially compact in
this space.
 Denote by
\begin{equation}
\label{010912}
( w^{(-)}, \hat y^{(-)}_{o},\hat y^{(+)}_{o})
\end{equation}
its $*$-weak limit.
Thanks to \eqref{032811a}, \eqref{b-ep} and the results of  Theorem \ref{cor011811} 
we conclude that
\begin{eqnarray}
\label{r-iota}
&&
\lim_{\eps\to0+}\ga a^{(\eps)}_{w,-} \delta w_{\eps}=0,\nonumber\\
&&
\lim_{\eps\to0+}\ga \left(\sum_{\iota=o,\pm}b^{(\eps)}_{\iota} \right)\delta w_{\eps}=0,\\
&&
\lim_{\eps\to0+}r_\eps^{(\pm)}=0.\nonumber
\end{eqnarray} 
Using Lemma \ref{lm011912}, equalities   \eqref{042008}
and the Lebesgue dominated convergence
theorem
we
conclude that
\begin{equation}
\label{1-2}
\lim_{\eps\to0+}a^{(\eps)}_{y,\iota}=-\frac{1}{2},\quad \iota\in\{-,+\}.
\end{equation}

Subtracting sideways from the first equation of \eqref{sys-1} the sum
of the remaining two and taking into account  \eqref{r-iota} and \eqref{1-2} we obtain 
\begin{equation}
\label{sys-1a1}
\lim_{\eps\to0+}\ga\left\{\left( a^{(\eps)}_{w} -2\ga\sum_{\iota=\pm} {\rm Re}\,b_{w,\iota}^{(\eps)}\right) w_{\eps}^{(-)}- \left( z_{\eps}^{(1,o)}-\frac{z_{\eps,o}^{(2,o)} }{2}\right)\right\}=0,
\end{equation}
where $z_{\eps,o}^{(2,o)} :=z_{\eps,o}^{(2,-)} +z_{\eps,o}^{(2,+)}$.
Moreover, a direct calculation shows that
\begin{eqnarray}
\label{010401}
&&
\ga\left( a^{(\eps)}_{w} -2\ga\sum_{\iota=\pm} {\rm Re}\,b_{w,\iota}^{(\eps)}\right) 
=\frac{4\ga}{3}\int_{\bbT}\frac{R\tilde f_\eps}{\tilde \delta_\eps}
dk,
\end{eqnarray}
where
\begin{eqnarray}
\label{tef}
&&
\tilde f_\eps:=\vphantom{\int_0^1}4\ga( R-R_\eps)\eta  \delta_\eps\om\bar\om
R' + 2\ga
      \eta ^2R''(k)(\ga R_\eps) \bar\om^2 \nonumber\\
      &&
+ 4\left[ \ga( R-R_\eps)(\ga R_\eps)( \delta_\eps\om )^2+ (\bar\om\delta_\eps\om )^2\right]\vphantom{\int_0^1}
-4(\ga R)\la\left[
 2\bar\om^2-2\left(\frac{\eps R' \eta }{2}\right)^2\right]\nonumber\\
&&
+
 4\la( \ga R_\eps)\left[ 4  \ga( R-R_\eps) (\ga R_\eps)+ 4 \bar\om^2 +
   (\eps\delta_\eps\om)^2- \left(\eps 
      R' \eta \right)^2  \right]
\\
&&
+ 4 \eps^{2} \left\{\frac{\la\ga}{2} R(\delta_\eps\om)^2+
\la^2 \left[5 \ga( R-R_\eps) (\ga
  R_\eps)+\bar\om^2+(\eps\delta_\eps\om)^2 
\right] \right\}
+ 8\eps^{4}\la ^3\ga R_\eps
+\eps^{6}\la^4.\nonumber
\end{eqnarray}
Using Lemma \ref{lm011912} and the Lebesgue dominated convergence theorem we obtain
\begin{equation}
\label{010301}
\lim_{\eps\to0+}\frac{4\ga}{3}\int_{\bbT}\frac{R\tilde f_\eps}{\tilde \delta_\eps}
dk=\frac{2\la}{3}+\frac{\eta^2}{3\ga}\int_{\bbT}\frac{(\om')^2}{R}dk.
\end{equation}
Using the above formula and substituting 
$$
z_{\eps}^{(1,o)}-\frac{z_{\eps,o}^{(2,o)}}{2}=\left(\frak{z}_\eps^{(-)}+\frak{z}_\eps^{(+)}\right)\cdot
{\rm e}
$$ 
(cf \eqref{032610} and \eqref{032610a}) we can rewrite   \eqref{sys-1a1} in the form
\begin{equation}
\label{sys-1a}
\left(\frac{2\la}{3}+\frac{\eta^2}{3\ga}\int_{\bbT}\frac{(\om')^2}{R}dk\right)w^{(-)}=\lim_{\eps\to0+}\sum_{j=1}^3V_{\eps,j},
\end{equation}
 where
\begin{eqnarray}
\label{032008}
&&
V_{\eps,1}=\frac{2\ga}{3}\int_{\bbT}\,{\frak w}^{(0)}_\eps(\la,
\eta,k)\cdot {\rm e}\,R dk,\nonumber
\\
&&
\\
&&
V_{\eps,2}=-\frac{2\ga (\pi \eta)^2}{3}\int_{\bbT} (\tilde \Delta_{1,\eps}-\tilde \Delta_{2,\eps})
\frac{R  f_{\eps}}{\tilde \delta_\eps}
dk,\nonumber\\
&&
\nonumber\\
&&
V_{\eps,3}:=\ga\eps\int_{\bbT}\tilde \Delta_{1,\eps}
\frac{R
 r_\eps^{(1)} }{\tilde\delta_\eps}dk
+\frac{\ga\eps}{2}\int_{\bbT}
(2\tilde d_{-}^{(\eps)}-\tilde d_{2}^{(\eps)}-\tilde d_{o}^{(\eps)}) \frac{R  r_\eps^{(2)} }{\tilde\delta_\eps} 
dk \nonumber\\
&&
 +\frac{\ga\eps}{2}\int_{\bbT}
[2(\tilde d_{+}^{(\eps)})^*-\tilde d_{o}^{(\eps)}-(\tilde d_{2}^{(\eps)})^*]\frac{ R
 r_{\eps,-}^{(2)} }{\tilde\delta_\eps}
dk. \nonumber
\end{eqnarray}
Here 
 \begin{equation}
\label{bwe}
({\frak w}^{(0)}_\eps)^T(\la, \eta,k):=[w_{\eps,+}^{(0)},y_{\eps,+}^{(0)},y_{\eps,-}^{(0)} ,w_{\eps,-}^{(0)} ]
\end{equation}
 is the solution of the system
\begin{equation}
\label{011812a}
\tilde D_\eps {\frak w}^{(0)}_\eps(\la, \eta,k)=\widehat{\frak  W}_\eps(\eta,k),
\end{equation}
where
$\widehat{\frak W}_\eps(\eta,k)$ is the column vector of Fourier-Wigner functions corresponding to the initial data, see   \eqref{032610-1}.
In addition,  $f_{\eps}$ is given by \eqref{f-eps} respectively, and $r_{\eps}^{(i)}$, $i=1,2$
satisfy \eqref{f}.

Thanks to Lemma \ref{lm011912} we conclude, upon an application of the
Lebesgue
dominated convergence theorem, that
\begin{equation}
\label{V-eps3}
\lim_{\eps\to0+}V_{\eps,3}=0.
\end{equation}
Using Theorems \ref{cor011811} and \ref{thm012610} and the definition
of $f_\eps$ (see \eqref{f-eps}) we get
\begin{equation}
\label{V-eps2b}
\lim_{\eps\to0+}V_{\eps,2}=-8\ga(\pi \eta)^2 w^{(-)}-2\ga(\pi \eta)^2
\bar v_{\phi},
\end{equation}
with (cf \eqref{wphi})
\begin{equation}
\label{barv}
\bar v_{\phi}(\la,\eta):=\bar w_\phi(\la,\eta)-\frac12\left(\bar y_{\phi,-}(\la,\eta) +\bar y_{\phi,+}(\la,\eta)
\right).
\end{equation}

\subsection{Limit of $V_{\eps,1}$}

Using the decomposition of the Wigner functions of the initial data into the parts corresponding to the 
macroscopic profile and the fluctuations, see \eqref{decomp}, we can
write an analogous decomposition
$
\widehat{\frak W}_\eps =\widehat{\overline{\frak W}}_\eps+\widehat{\widetilde{\frak W}}_\eps,
$
 for the Laplace-Fourier transforms of the respective Wigner functions.
It allows us to write
$
 V_{\eps,1}= V_{\eps,1}^{(1)} +V_{\eps,1}^{(2)},
 $
 where 
\begin{eqnarray*}
&&
V_{\eps,1}^{(1)}:=\frac{2\ga}{3}\int_{\bbT}\,\overline{\frak w}^{(0)}_\eps(\la,
\eta,k)\cdot {\rm e}\,R dk,\\
&&
V_{\eps,1}^{(2)}:=\frac{2\ga}{3}\int_{\bbT}\,\widetilde{\frak w}^{(0)}_\eps(\la,
\eta,k)\cdot {\rm e}\,R dk.
\end{eqnarray*}
Here
$$
\overline{\frak w}^{(0)}_\eps=[\overline w_{\eps,+}^{(0)},\overline
y_{\eps,+}^{(0)},\overline y_{\eps,-}^{(0)},\overline
w_{\eps,-}^{(0)}], \quad 
\widetilde{\frak w}^{(0)}_\eps=[\widetilde w_{\eps,+}^{(0)},\widetilde 
y_{\eps,+}^{(0)},\widetilde  y_{\eps,-}^{(0)},\widetilde 
w_{\eps,-}^{(0)}]
$$
are the solutions of the analogues of \eqref{011812a} in which the right
hand side has been replaced by $\widehat{\overline{\frak W}}_\eps$ and
$\widehat{\widetilde{\frak W}}_\eps$, respectively.

\subsubsection{Macroscopic
Wigner functions and their dynamics}
From \eqref{010801} we get
\begin{equation}
\label{091308}
\partial_t\hat\phi(t,k)=-i\tau_2(\pi k)^2\hat \phi (t,k) -6\gamma\pi^2 k^2\left(\hat \phi(t,k)-\hat \phi^*(t,-k)\right).
\end{equation} 
Therefore the Fourier transforms $\widehat{\frak W} _{\phi}(t)$  of the macroscopic
Wigner functions (cf \eqref{011308})
satisfy
\begin{eqnarray}
\label{011008}
&&
\partial_t\widehat{W} _{\phi,+}=-2\pi^2\left\{i\tau_2k\eta
+6\gamma
\left[k^2+\left(\frac{\eta}{2}\right)^2\right]\right\}\widehat{W}_{\phi,+} 
+6\gamma\pi^2\sum_{\iota'\in\{-,+\}}\left(k-\iota'\frac{
    \eta}{2}\right)^2\widehat{Y}_{\phi,\iota'} ,\nonumber\\
&&
\\
&&
\partial_t\widehat{Y}_{\phi,+}=-2\pi^2\left(i\iota\tau_2
+6\gamma
\right)\left[k^2+\left(\frac{\eta}{2}\right)^2\right]\widehat{Y}_{\phi,+}
+6\gamma\pi^2\sum_{\iota'=\pm1}\left(k-\iota'\frac{
    \eta}{2}\right)^2\widehat{W}_{\phi,\iota'} .\nonumber
\end{eqnarray}
Taking the Laplace transforms of both sides of \eqref{011008} we
obtain
\begin{equation}
\label{020911aa}
\tilde D_{1} w_{\phi,+} +\tilde D_{+} y_{\phi,+} +\tilde D_{-}  y_{\phi,-}
=\widehat{W}_{\phi,+} (0,\eta,k)
\end{equation}
and
\begin{equation}
\label{040911aa}
\tilde D_{+} w_{\phi,+}+\tilde D_{2} y_{\phi,+} +\tilde D_{-}
w_{\phi,-}=\widehat{Y}_{\phi,+} (0,\eta,k),
\end{equation}
where
\begin{eqnarray}
\label{030911aa}
&&
\tilde D_{1}:= \la+2\pi^2\left\{6\gamma
\left[k^2+\left(\frac{\eta}{2}\right)^2\right]+i\tau_2 k\eta\right\},\\
&&
\tilde D_{2}:=\la+2\pi^2 \left[k^2+\left(\frac{\eta}{2}\right)^2\right]( 6\gamma
+i\tau_2),\nonumber\\
&&
\tilde  D_{\pm}:=-6\gamma\pi^2\left(k\mp\frac{
    \eta}{2}\right)^2.
\end{eqnarray}
An elementary calculation shows that
\begin{equation}
\label{030911a}
\lim_{\eps\to0+}\eps^{-2}\tilde D_{j}^{(\eps)}(\la,q,\eps k)= \tilde
D_{j}(\la,\eta, k),\quad j\in\{1,2,-,+\}
\end{equation}
for any $\la>0$ and $(\eta,k)\in\bbR^2$.

The closed system of linear algebraic equations for the components of
the  Laplace-Fourier transforms ${\frak w} _{\phi}$ (cf
\eqref{032810}) takes the form
\begin{equation}
\label{0118121aa}
\tilde D {\frak w}_\phi=\widehat{\frak W}_\phi(0),
\end{equation}
where
\begin{equation}
 \label{011612aab}
 \tilde D=
 \left[
 \begin{array}{cc}
A &B \\
B&C
 \end{array}
 \right]
  \end{equation}
with
\begin{equation}
 \label{011612aac}
 A=
 \left[
 \begin{array}{cc}
\tilde D_{1}& \tilde D_{+}  \\
\tilde D_{+}& \tilde D_{2}
 \end{array}
 \right],\qquad C=
 \left[
 \begin{array}{cc}
 \tilde D_{2}^* &\tilde D_{+} \\
\tilde D_{+} &\tilde D_{1}^*
 \end{array}
 \right]
  \end{equation}
and
$
B=\tilde D_{-}I_2
$ (cf \eqref{011612} and \eqref{012410}). It can be checked by a
direct inspection that $[A,B]=[B,C]=[A,C]=0$.
Therefore, 
\begin{eqnarray}
\label{013110}
&&
\tilde\delta(\la,\eta,k):={\rm det}\tilde D(\la,\eta,k)={\rm det}(AC-B^2)\\
&&
={\rm det}\tilde D(\la,q,k)=|\tilde D_{1}\tilde D_{2}^*+\tilde D_{+}^2-\tilde D_{-}^2 |^2-4\tilde D_{+}^2{\rm Re}\,
\tilde D_{1}\,{\rm Re}\,
\tilde D_{2}.\nonumber
\end{eqnarray}
Thanks to \eqref{030911a} we conclude that 
\begin{eqnarray}
\label{010907a}
&&
\lim_{\eps\to0+}\eps^{-8}\tilde\delta_\eps(\la,\eta,\eps k)=\tilde\delta (\la, \eta, k)
\end{eqnarray}
for each $\la>0$ and $(\eta,k)\in\bbR^2$. 

The above, combined with \eqref{051612}, implies that
for any $M>0$ and $\la_0$ as in Proposition \ref{023112} we have
 \begin{equation}
\label{051612b}
\tilde \delta(\la,\eta,k)>0,\quad |\eta|<M,\,k\in\bbR,\,\la>\la_0.
\end{equation}
The matrix $\tilde D(\la,\eta,k)$ is then invertible and, cf \eqref{011208}, $\tilde D^{-1}=
\tilde\delta^{-1}{\rm adj}(\tilde D)$. The adjugate of
$\tilde D$ equals
\begin{equation}
\label{011208aa}
{\rm adj}(\tilde D )=
\left[\begin{array}{ll}
 P & Q \\
 Q &M 
 \end{array}
 \right], 
\end{equation}
where $M $, $P $ and  $Q $ are $2\times 2$ matrices given by
\begin{eqnarray*}
 &&P :=\left[\begin{array}{ll}
 \tilde d_{1} &\tilde d_{-} \\
 \tilde d_{-} &\tilde d_{2} 
 \end{array}
 \right],\quad
Q :=\left[\begin{array}{ll}
 (\tilde d_{+} )^*&\tilde d_{o} \\
 \tilde  d_{o} &\tilde d_{+} 
 \end{array}
 \right],\\
&&
M :=
\left[\begin{array}{ll}
(\tilde  d_{2} )^*&(\tilde d_{-} )^*\\
(\tilde d_{-} )^*& (\tilde d_{1} )^*
 \end{array}
 \right].
\end{eqnarray*}
Here
\begin{eqnarray}
\label{td}
&&
\tilde d_1:=|\tilde D_2|^2 \tilde D_1^*-(\tilde D_+^2+ \tilde
D_-^2){\rm Re}\,\tilde D_2-i (\tilde D_+^2- \tilde
D_-^2) {\rm Im}\,\tilde D_2,\nonumber\\
&&
\tilde d_2:=|\tilde D_1|^2 \tilde D_2^*-(\tilde D_+^2+ \tilde
D_-^2){\rm Re}\,\tilde D_1-i (\tilde D_+^2- \tilde
D_-^2) {\rm Im}\,\tilde D_1,\nonumber\\
&&
\tilde d_-:=\tilde D_+ (\tilde D_+^2- \tilde
D_-^2)- \tilde D_+\tilde D_1^*{\rm Re}\,\tilde D_2+i \tilde
D_1^*\tilde D_+ {\rm Im}\,\tilde D_2 \nonumber\\
&&
=\tilde D_+ (\tilde D_+^2- \tilde
D_-^2)- \tilde D_+\tilde D_2^*{\rm Re}\,\tilde D_1+i \tilde
D_2^*\tilde D_+ {\rm Im}\,\tilde D_1,\nonumber\\
&&
\tilde d_{+}:=-\tilde D_{-}(\tilde D_{1}\tilde D_{2}^*+\tilde
D_{+}^2-\tilde D_{-}^2),\nonumber\\
&&
\tilde d_{o}:= 2\tilde D_{+}\tilde D_{-}{\rm Re}\,
\tilde D_{2}.
\end{eqnarray}
Thanks to \eqref{030911a} we conclude that 
\begin{eqnarray}
\label{010907}
&&
\lim_{\eps\to0+}\eps^{-6}\tilde d_{j}^{(\eps)}(\la, \eta,\eps k)= \tilde
d_{j}(\la, \eta, k),\quad j\in\{-,+,1,2,o\}
\end{eqnarray}
for each $\la>0$ and $(\eta,k)\in\bbR^2$. 

\bigskip

\subsubsection{Limit of $V_{\eps,1}^{(1)}$}
The limit in question is a special case of the following result.
\begin{prop}
\label{prop011508}
Suppose that $\varphi\in C^2(\bbT)$ is such that
$
\varphi (0)=\varphi '(0)= 0.
$
Then, for any $M>0$  we have
\begin{equation}
\label{011708}
\lim_{\eps\to0+}\int_{\bbT}\overline w_{\eps,\pm}^{(0)}(\la,
\eta,k) \varphi (k) dk=\frac12 \varphi ''(0)\int_{\bbR} k^2w_{\phi,\pm}(\la,
\eta,k) dk
\end{equation}
and
\begin{equation}
\label{021708}
\lim_{\eps\to0+}\int_{\bbT}\overline y_{\eps,\pm}^{(0)}(\la,
\eta,k) \varphi (k) dk=\frac12 \varphi ''(0)\int_{\bbR} k^2y_{\phi,\pm}(\la,
\eta,k) dk,\quad |\eta|\le M,\quad \la>\la_0.
\end{equation}
Here $w_{\phi,\pm} $ and $y_{\phi,\pm}$ are given by \eqref{022810}.
\end{prop}
\proof
We only prove \eqref{011708}, as the argument for \eqref{021708} is very similar.
The left hand side of \eqref{011708} for $\overline w_{\eps,+}^{(0)}$ can be rewritten in the form
$$
\int_\bbT\tilde \delta_\eps^{-1}\left[
 \tilde d_{1}^{(\eps)}\widehat{\overline{ W}}_{\eps,+}+\tilde d_{-}^{(\eps)}\widehat{\overline{ Y}}_{\eps,+}+(\tilde d_{+}^{(\eps)})^*\widehat{\overline{ Y}}_{\eps,-}+
 \tilde d_{o}^{(\eps)}\widehat{\overline{ W}}_{\eps,-}\right]\varphi(k) dk.
$$
Denote by $J_{j,\eps}$, $j=1,2,3,4$  the respective terms arising after
opening of the square bracket. Changing variables $k:=k/\eps$ we can
write (cf \eqref{041308})
$$
J_{1,\eps}=\frac12\sum_{x,x'}\int_{-1/(2\eps)}^{1/(2\eps)} (\tilde
\delta_\eps^{-1}\tilde d_{1}^{(\eps)})(\la,\eta,\eps k)\hat\phi^*\left(k+\frac{x}{\eps}-\frac{\eta}{2}\right)\hat\phi\left(k+\frac{x'}{\eps}+\frac{\eta}{2}\right) \varphi (\eps k) dk.
$$
Thanks to Lemma \ref{lm011912} we conclude that there exist $\la_0,\eps_0>0$
such that for any $\la>\la_0$ we have
$$
\sup_{\eps\in(0,\eps_0]}\sup_{k,|\eta|\le M}R(\eps k)\left|\frac{\tilde d_{\iota}^{(\eps)}(\la, \eta,\eps k)}{\tilde\delta_\eps(\la,\eta,\eps k)}\right|<+\infty.
$$
In addition, we have
$$
\sup_{\eps\in(0,1]}\sup_{|k|\le 1/(2\eps)}\frac{(\eps |k|)^2}{R(\eps k)}<+\infty.
$$
Therefore,
$$
\sup_{\eps\in(0,\eps_0]}\sup_{k,|\eta|\le M}|\varphi(\eps k)|\left|\frac{\tilde d_{\iota}^{(\eps)}(\la, \eta,\eps k)}{\tilde\delta_\eps(\la,\eta,\eps k)}\right|<+\infty.
$$
In fact, thanks to the rapid decay of the macroscopic wave function $\phi$,  we can write
\begin{equation}
\label{031708}
\lim_{\eps\to0+}J_{1,\eps}=\frac14\lim_{\eps\to0+}\int_{-1/(2\eps)}^{1/(2\eps)} (\tilde \delta_\eps^{-1} \tilde d_{1}^{(\eps)})(\la,\eta,\eps k)\hat\phi^*\left(k-\frac{\eta}{2}\right)\hat\phi\left(k+\frac{\eta}{2}\right)\eps^2(\varphi  ''(0) k^2+o(1)) dk.
\end{equation}
 By virtue of the Lebesgue dominated convergence theorem, we conclude that the limit in \eqref{031708}
equals
$$
\frac{\varphi''(0)}{4}\int_{\bbR} (\tilde \delta^{-1}\tilde d_{1})(\la,\eta, k)k^2\hat\phi^*\left(k-\frac{\eta}{2}\right)\hat\phi\left(k+\frac{\eta}{2}\right)dk.
$$
Dealing similarly with the remaining terms $J_{j,\eps}$, $j=2,3,4$ we
conclude \eqref{011708} for $\overline w_{\eps,+}^{(0)}$. The cases of
 $\overline w_{\eps,-}^{(0)}$ and $\overline y_{\eps,\pm}^{(0)}$ can
 be handled similarly. 
\qed

\bigskip

Since $R(0)=R'(0)=0$ and $R''(0)=12\pi^2$ (cf \eqref{beta1} and \eqref{beta2}),  by a direct application of Proposition \ref{prop011508},
we obtain 
\begin{equation}
\label{V-eps1b}
\lim_{\eps\to0+}V_{\eps,1}^{(1)}=8\ga\pi^2\int_{\bbR}k^2\frak
w_\phi(\la,\eta,k)\cdot {\rm e}\,dk
\end{equation}
for all $|\eta|\le M$ and $\la>\la_0$.

\bigskip

\subsubsection{Limit of $V_{\eps,1}^{(2)}$}

For any $J\in {\cal S}(\bbR)$ such that $\hat J$ is 
supported in $[-M,M]$ we can write
\begin{eqnarray}
\label{011007a}
&&\int_\bbR V_{\eps,1}^{(2)}(\la,\eta)\hat J^*(\eta)d\eta\\
&&
=
\frac{2\ga}{3}\int_\bbR\int_{\bbT}\hat J^*\{\widehat{\widetilde W}_{\eps,+}\tilde \Delta_{1,\eps}+\widehat{\widetilde Y}_{\eps,+}\tilde \Delta_{2,\eps}+\widehat{\widetilde Y}_{\eps,-} \tilde \Delta_{2,\eps}^*+
\widehat{\widetilde W}_{\eps,-}\tilde \Delta_{1,\eps}^*\}\frac{Rd\eta dk}{\tilde \delta_{\eps}}.\nonumber
\end{eqnarray}
By virtue of Lemma \ref{lm011912} we can use the Lebesgue dominated
convergence theorem to  enter with the limit, as
$\eps\to0+$, under the integral.

Combining \eqref{042910} and \eqref{052910} 
we
conclude  that
\begin{eqnarray}
\label{V-eps1c}
&&\lim_{\eps\to0+}\int_\bbR V_{\eps,1}^{(2)}\hat J^*(\eta)d\eta=\frac{2}{3}\lim_{\eps\to0+}\int_\bbR\int_{\bbT}\hat J^*(\eta)\widehat{\widetilde
  W}_{\eps,+}(\eta,k)d\eta dk \nonumber\\
&&
=\lim_{\eps\to0+}\frac{\eps}{3}\sum_x|\tilde
\psi_x^{(\eps)}|^2J(\eps x)=
\frac{2}{3}\int_{\bbR}e_{\rm th}(y)J^*(y)dy.
\end{eqnarray}
The penultimate equality follows from \eqref{eq:waveenergy}.

\section{Proof of Theorem \ref{thm012610}}

\label{sec4a}

\subsection{Proof of $(\ref{010707})$}

\label{sec15.4}

Recall that $\tilde
d_{j}(\la, \eta, k)$, $j\in\{1,2,o,-,+\} $ and $  {\rm det}\,\tilde D(\la,\eta,k)$ are
given by \eqref{td} and \eqref{013110} respectively. 
We recall also 
$\tilde\Delta_j(\la,q,k)$, $j=1,2$ are defined by a modification of  formulas
\eqref{D-1} where the coefficients $ \tilde d_{\iota}^{(\eps)}$ have
been replaced by the corresponding $\tilde
d_{\iota}$.

Given $\varphi\in C(\bbT)$ we can write
$$
\int_{\bbT}  w_{\eps,+}(\la,\eta,k)\varphi(k)dk=
\int_{\bbT} (I_\eps+I\!I_\eps+I\!I\!I_\eps+I\!V_\eps)\varphi(k)dk.
$$
Here  $I_\eps$, $I\!I_\eps$, $I\!I\!I_\eps$, $I\!V_\eps$ are
given by \eqref{w-eps}.
By virtue of \eqref{021112} we conclude that
$$
\lim_{\eps\to0+}\int_{\bbT} I_\eps\varphi(k)dk=\lim_{\eps\to0+} w_\eps^{(-)}\int_{\bbT} \varphi(k)dk,\quad |\eta|\le M,\,\la>\la_0.
$$
\begin{lm}
\label{lm010907}
For any $M>0$ 
\begin{equation}
\label{030907}
\lim_{\eps\to0+}\int_{\bbT} |I\!V_\eps|dk=0,\quad |\eta|\le M,\,\la>\la_0.
\end{equation}
\end{lm}
\proof
Note that, according to Proposition \ref{023112}, for each $M>0$ we can
choose $\la_0,\eps_0>0$ such that for $\la>\la_0$ 
\begin{eqnarray}
\label{020907-1}
&&\int_{\bbT} |I\!V_\eps(\la,p,k)|dk\le \eps^{3}\left(\sum_{j=1}^4 \|r_\eps^{(j)}\|_{{\cal
    A}'}\right)\sup_k\tilde \delta_{\eps}^{-1}\left(\sum_j|\tilde d_j^{(\eps)}|\right)\nonumber\\
&&
\\
&&
\preceq \eps^{3}\left(\sum_{j=1}^4 \| r_\eps^{(j)}\|_{{\cal
    A}'}\right)\sup_k\left(\tilde\delta_\eps^{(0)}\right)^{-1}\left(\sum_j|\tilde d_j^{(\eps)}|\right)\nonumber
\end{eqnarray}
for $|\eta|\le M$ and $\eps\in(0,\eps_0]$.
Thanks to \eqref{012911aax1} we conclude that
\begin{equation}
\label{050907}
\eps^{3}\left(\tilde\delta_\eps^{(0)}\right)^{-1}|\tilde
d_-^{(\eps)}|
\preceq
\eps^{3} \left(\tilde\delta_\eps^{(0)}\right)^{-1}\left(\sum_{j=0}^3R_\eps^{3-j}(\la\eps^{2})^j \right).
\end{equation}
Invoking the definition of $\tilde\delta_\eps^{(0)}$, see \eqref{082312}, we can bound the right hand side of
\eqref{050907} by
$
\eps^{3}
\left(R_\eps+\la\eps^2
\right)^{-1}\preceq \eps.
$
The conclusion of the lemma follows then directly from the above
estimate and \eqref{f}.
\qed

\bigskip

Using a similar argument
we infer that for any $\la>\la_0$, $|\eta|\le M$ we have
\begin{eqnarray}
\label{020907}
&&\sup_k| I\!I\!I_\eps|\preceq \eps^{2}\sup_k\tilde \delta_{\eps}^{-1}\left(\sum_j|\tilde d_j^{(\eps)}|\right)\nonumber\\
&&
\\
&&
\preceq \eps^{2}\sup_k\left(\tilde\delta_\eps^{(0)}\right)^{-1}\left(\sum_j|\tilde d_j^{(\eps)}|\right)\preceq 1\nonumber
\end{eqnarray}
for $\eps\in(0,\eps_0]$, thanks to  \eqref{012911aax1} and
\eqref{082312}.
On the other hand, due to Proposition \ref{lm012008}, for any $\la>\la_0$, $k\not=0$ and $|\eta|\le M$ we have
\begin{equation}
\label{023110}
\lim_{\eps\to0+}\eps^{2}\tilde
\delta_{\eps}^{-1}(\la,\eta,k)\left(\sum_j|\tilde d_j^{(\eps)}(\la,\eta,k)|\right)=0
\end{equation}
By virtue of the Lebesgue dominated convergence theorem we conclude
therefore that
$$
\lim_{\eps\to0+}\int_{\bbT} I\!I\!I_\eps\varphi(k)dk=0.
$$

\bigskip

Finally, we have
\begin{equation}
\label{011007}
\int_{\bbT} I\!I_\eps\varphi(k)dk=
\eps^{2}\int_{\bbT}\tilde \delta_{\eps}^{-1}\{\widehat W_{\eps,+}\tilde d_1^{(\eps)}+\widehat Y_{\eps,+}\tilde d_{-}^{(\eps)}+\widehat Y_{\eps,-} (\tilde d_{+}^{(\eps)})^*+
\widehat W_{\eps,-}\tilde d_{o}^{(\eps)}\}\varphi(k)dk.
\end{equation}
The computation of the limit, as $\eps\to0+$, of each of the four
 expressions 
$J_j^{(\eps)}$, $j=1,\ldots,4$ that arise
in the right hand side after opening of the bracket is almost identical so we
 explain only how to deal with the first one.
 Using \eqref{041308} we can write that 
$
J_1^{(\eps)}=\sum_{j=1}^2J_{1j}^{(\eps)},
$
with (cf \eqref{041308})
\begin{eqnarray*}
&&
J_{11}^{(\eps)}:=\eps^2 \int_{\bbT}
  \delta_{\eps}^{-1}\tilde
  d_1^{(\eps)}\widehat{ \overline{W}}_{\eps,+}(\eta,k)\varphi(k)dk,\\
&&
J_{12}^{(\eps)}:=\eps^2 \int_{\bbT}\left(\tilde
  \delta_{\eps}^{-1}\tilde
  d_1^{(\eps)}\right)\widehat{ \widetilde{W}}_{\eps,+}(\eta,k)\varphi(k)dk.
\end{eqnarray*}
In what follows we show that
\begin{equation}
\label{061308a}
\lim_{\eps\to0+}J_{12}^{(\eps)}=0.
\end{equation}
and, cf \eqref{011308}, 
\begin{equation}
\label{061308b}
\lim_{\eps\to0+}J_1^{(\eps)}=\lim_{\eps\to0+}J_{11}^{(\eps)}=\varphi(0)\int_{\bbR}
\frac{\widehat{W}_{\phi,+}(0) \tilde d_1}{{\rm det}\,\tilde D}dk.
\end{equation}
We repeat the above argument to compute the limits of the remaining
terms $J_j^{(\eps)}$ and obtain  that, cf \eqref{0118121aa}
\begin{eqnarray}
\label{021007}
&&\lim_{\eps\to0+}\int_{\bbT} I\!I_\eps\varphi(k)dk\\
&&
=\varphi(0)\int_{\bbR}{\rm det}\,\tilde D^{-1}\{\widehat{ W} _{\phi,+}(0)\tilde d_1+\widehat{ Y} _{\phi,+}(0)\tilde d_-+\widehat{ Y} _{\phi,-}(0)(\tilde d_{+})^*+
\widehat{ W} _{\phi,-}(0)\tilde d_{o}\}dk\nonumber\\
&&
=\varphi(0) \int_{\bbR}w _{\phi,+}(\la,\eta,k)dk.\nonumber
\end{eqnarray}



\subsubsection{Proof of (\ref{061308b})}

After the change of variables $k':= k/\eps$ we can write
\begin{equation}
\label{051308-1}
J_1^{(\eps)}=\sum_{x,x'}\int_{-1/(2\eps)}^{1/(2\eps)}
\frac{\eps^{-6}\tilde d_1^{(\eps)}(\la,\eta,\eps k)}{\eps^{-8}\tilde \delta_{\eps}(\la,\eta,\eps k)}
\hat\phi^*\left(\frac{x}{\eps}+k-\frac{\eta}{2}\right)\hat\phi\left(\frac{x'}{\eps}+k+\frac{\eta}{2}\right)\varphi(\eps k)dk.
\end{equation}
Using the argument from  the proof of 
Lemma \ref{lm010907} we conclude that for any $M>0$ and $\la>\la_0$, where
$\la_0,\eps_0>0$ are as in the statement of Proposition \ref{023112}, 
 \begin{equation}
\label{051308}
\left|\frac{\eps^{-6}\tilde d_1^{(\eps)}(\la,\eta ,\eps k)}{\eps^{-8}\tilde \delta_{\eps}(\la,\eta ,\eps k)}\right|\preceq 1,
\end{equation}
for all $k\in\bbR$, $|\eta|\le M$ and $\eps\in(0,\eps_0]$. 
Due to the  decay of the wave function $\hat\phi$ we conclude that
$$
\lim_{\eps\to0+}J_1^{(\eps)}=\lim_{\eps\to0+}\int_{-1/(2\eps)}^{1/(2\eps)}
\frac{\eps^{-6}\tilde d_1^{(\eps)}(\la,\eta ,\eps k)}{\eps^{-8}\tilde \delta_{\eps}(\la,\eta ,\eps k)}
\hat\phi^*\left(k-\frac{\eta }{2}\right)\hat\phi\left(k+\frac{\eta }{2}\right)\varphi(\eps k)dk.
$$
Thanks to \eqref{051308} to compute the last limit we can use the Lebesgue dominated convergence
and conclude, using \eqref{010907a} and \eqref{010907}, that 
the right hand side of  the above equality coincides with the right
hand side of \eqref{061308b}.

\subsubsection{Proof of (\ref{061308a})}

Using condition
\eqref{finite-energy1} we conclude that for some $r>1$
\begin{equation}
\label{033110}
\sup_{\eta}\int_{\bbT}|\widehat{ \widetilde{W}}_{\eps,+}(\eta,k)|^rdk<+\infty.
\end{equation}
Combining the above with estimate \eqref{020907} together with the
limit \eqref{023110} we conclude that
for any  $\la>\la_0$ and
$|\eta|\le M$ 
$$
\lim_{\eps\to0+}\eps^2 \int_{\bbT}|\widehat{ \widetilde{W}}_{\eps,+}(\eta,k)\varphi(k)| \left|\frac{\tilde
  d_1^{(\eps)}(\la,\eta,k)}{\tilde
  \delta_{\eps}(\la,\eta,k)}\right|dk=0
$$
This obviously implies \eqref{061308a}.


\label{sec8.4}

\bigskip

\subsection{ Proof of $(\ref{031811b})$} 

We use the notation from  Section \ref{sec8.5} and  carry
out our analysis only for $y_{\eps,+}$, as the argument for
$y_{\eps,-}$ is very similar. For any $\varphi\in C(\bbT)$ we have
\begin{equation}
\label{071308}
\int_{\bbT} y_{\eps}(\la,q,k)\varphi dk=
\int_{\bbT} I_\eps\varphi dk+ \int_{\bbT} I\!I_\eps\varphi dk+
\int_{\bbT} I\!I\!I_\eps\varphi dk
+\int_{\bbT} I\!V_\eps\varphi dk.
\end{equation}
The analysis of the terms on the right hand side of \eqref{071308} is very similar to the one done in Section \ref{sec8.4}. As a result we obtain
$$
\lim_{\eps\to0+}\int_{\bbT}(|I_\eps|+| I\!I\!I_\eps |+| I\!V_\eps|)dk=0.
$$
In addition,
\begin{eqnarray}
\label{021007a}
&&
\lim_{\eps\to0+}\int_{\bbT} I\!I_\eps dk\\
&&
=\varphi(0)\int_{\bbR}{\rm det}\,\tilde D^{-1}\{\widehat{ W}_{\phi,+}\tilde d_-+\widehat{  Y}_{\phi,+}\tilde d_2+\widehat{ Y}_{\phi,-}(\tilde d_{o})^*+
\widehat{  W}_{\phi,-}\tilde d_{-}\}dk\nonumber\\
&&
=\varphi(0)\int_{\bbR} y_{\phi,+}(\la,\eta,k)dk\nonumber
\end{eqnarray}
and (\ref{031811b})
follows.

\section{Proof of Theorem \ref{profiles}}

\label{sec-p-k}

Suppose that $\kappa(t,y)$ and $p(t,y)$ satisfy \eqref{010801}.
Then, $\hat \phi(t,k)$ -- the Fourier transform of 
$$
\phi(t,y):=\frac{\tau_2}{4}\kappa(t,y)+ip(t,y)
$$
satisfies
 \begin{equation}
 \label{basic:sde:2ab}
\frac{ d}{dt}\hat \phi(t,k)=-i\tau_2(\pi
k)^2\hat \phi(t,k)-6\ga \pi^2k^2
  \left[\hat \phi(t,k)-(\hat \phi)^*(t,-k)\right].
 \end{equation}
Let
$$
\hat{\bar \psi}_\eps(t,k):=\eps \bbE_\eps \hat\psi^{(\eps)}(t,\eps k).
$$
From \eqref{011307a} we obtain
  \begin{equation}
 \label{basic:sde:2a}
\frac{ d}{dt}\hat{\bar \psi}_{\eps} (t,k)=\frac{-i\om(\eps
  k)}{\eps^{2}}\hat{\bar \psi}_{\eps}(t,k)-\frac{\ga
   R(\eps k)}{\eps^{2}}\left[\hat{\bar \psi}_{\eps}(t,k)-(\hat{\bar \psi}_{\eps})^*(t,-k)\right].
 \end{equation}
After a straightforward calculation we obtain, using \eqref{limits0} that   for any $G\in
C_0^\infty(\bbR)$
\begin{equation}
\label{010111}
\int_{\bbR}G(y) \phi(y)dy=\lim_{\eps\to0+}\eps\sum_xG(\eps x)\langle {\psi}_x\rangle_{\mu_\eps}
=
\lim_{\eps\to0+}\int_{\bbR}\hat G(k) \hat{\bar{\psi}}_\eps\left(0,- k\right)dk.
\end{equation}
Since
$$
\lim_{\eps\to0+}\frac{\om(\eps
  k)}{\eps^{2}}=\tau_2\pi^2 k^2\quad\mbox{and}\quad \frac{
   R(\eps k)}{\eps^{2}}=6\pi^2k^2,
$$
uniformly on compact intervals,
an elementary stability theory for solutions of ordinary differential equations guarantees
that for any $T,M>0$ we have
\begin{equation}
\label{031703}
\hat{\bar \psi}_{\eps}(t,k)=(1+o(1))\hat{\bar \psi}_\eps^{(0)}(t,k),
\end{equation}
uniformly on  $|k|\le M,\,|t|\le T$,  as $\eps\ll1$,
where $\hat{\bar \psi}_\eps^{(0)}(0,k)$ satisfies 
\eqref{basic:sde:2ab}
with the initial condition
$
\hat{\bar \psi}_\eps^{(0)}(0,k):=\hat{\bar \psi}_\eps(0,k).
$
Equation \eqref{basic:sde:2ab} can be solved explicitly. 
Taking into account \eqref{010111} we obtain, upon letting
$\eps\to0+$,
that
\begin{equation}
\label{010111a}
\lim_{\eps\to0+}\int_{\bbR}\hat G(k) \hat{\bar{\psi}}_\eps^{(0)}\left(t,- k\right)dk=\int_{\bbR}\hat G(k) \hat{\bar\psi}\left(t,- k\right)dk,
\end{equation}
where  $\hat{\bar\psi}\left(t,k\right)$ satisfies
\eqref{basic:sde:2ab} with the initial
$
\hat{\bar \psi}(0,k):=\hat{ \phi}(0,k).
$
Therefore $
\hat{\bar \psi}(t,k)=\hat{ \phi}(t,k)
$ and, in conclusion,
\begin{eqnarray*}
&&\lim_{\eps\to0+}\eps\sum_xG(\eps x)\bbE_\eps {\psi}_x^{(\eps)}(t) =
\lim_{\eps\to0+}\int_{\bbR}\hat G(k) \hat{\bar{\psi}}_\eps\left(t,- k\right)dk
\\
&&=\int_{\bbR}\hat G(k) \hat{{\phi}}\left(t,- k\right)dk= \int_{\bbR}
G(y) {\phi}\left(t,y\right)dy.
\end{eqnarray*}
\qed

\end{document}